\documentclass[lettersize,journal]{IEEEtran}
\usepackage{amsmath,amsfonts}
\usepackage{algorithmic}
\usepackage{algorithm}
\usepackage{array}
\usepackage[caption=false,font=normalsize,labelfont=sf,textfont=sf]{subfig}
\usepackage{textcomp}
\usepackage{stfloats}
\usepackage{url}
\usepackage{verbatim}
\usepackage{graphicx}
\usepackage{cite}
\usepackage{latexsym}
\usepackage{amsbsy}
\usepackage{cite}
\usepackage{hyperref}
\usepackage{esint}
\usepackage{tikz}
\usepackage{graphbox}
\usepackage{graphicx}
\usepackage{caption}
\newsavebox{\imagebox}
\usepackage{makecell}
\usepackage{multirow}
\usepackage{longtable}
\usepackage{url}
\usepackage{algorithm,algorithmic}
\usepackage{mathtools}
\usepackage[switch]{lineno}
\usepackage{booktabs}
\usepackage{dsfont}
\usepackage{mathrsfs}
\usepackage[scr=boondoxo]{mathalfa}
\usepackage{bbm}
\usepackage{multirow}

                    {$\blacksquare$\vspace*{7pt}} 

\def\b{\mathbf{b}}
\def\x{\mathbf{x}}
\def\y{\mathbf{y}}

\def\u{\mathbf{u}}
\def\v{\mathbf{v}}
\def\w{\mathbf{w}}

\def\R{{\mathbb R}}

\def\f{\frac}

\def\a{{\boldsymbol a}}

\def\bi{{\mathbf i}}

\def\n{\mathbf{n}}

\def\bi{\begin{itemize}} \def\ei{\end{itemize}}
\def\be{\begin{eqnarray*}}
\def\ee{\end{eqnarray*}}

\def\0{{\mathbf 0}}

\newcommand{\beq}{\begin{equation}}
\newcommand{\eeq}{\end{equation}}

\def\XXint#1#2#3{{\setbox0=\hbox{$#1{#2#3}{\int}$ }
\vcenter{\hbox{$#2#3$ }}\kern-.55\wd0}}

\begin{document}

\title{Fully automatic integration of dental CBCT images and full-arch intraoral impressions with stitching error correction via individual tooth segmentation and identification}

\author{Tae~Jun~Jang,~Hye~Sun~Yun,~Chang~Min~Hyun,~Jong-Eun~Kim,~Sang-Hwy~Lee,~and~Jin~Keun~Seo \thanks{Tae Jun Jang, Hye Sun Yun, Chang Min Hyun, and Jin Keun Seo are with the School of Mathematics and Computing (Computational Science and Engineering), Yonsei University, Seoul, South Korea. \protect (Email: sally7711548@yonsei.ac.kr (Corresponding author))} \thanks{Jong-Eun Kim is with the Department of Prosthodontics, College of Dentistry, Yonsei University, Seoul, South Korea.} \thanks{Sang-Hwy Lee is with the Department of Oral and Maxillofacial Surgery, Oral Science Research Center, College of Dentistry, Yonsei University, Seoul, South Korea.}}



\maketitle

\begin{abstract}
We present a fully automated method of integrating intraoral scan (IOS) and dental cone-beam computerized tomography (CBCT) images into one image by complementing each image's weaknesses. Dental CBCT alone may not be able to delineate precise details of the tooth surface due to limited image resolution and various CBCT artifacts, including metal-induced artifacts. IOS is very accurate for the scanning of narrow areas, but it produces cumulative stitching errors during full-arch scanning. The proposed method is intended not only to compensate the low-quality of CBCT-derived tooth surfaces with IOS, but also to correct the cumulative stitching errors of IOS across the entire dental arch. Moreover, the integration provide both gingival structure of IOS and tooth roots of CBCT in one image. The proposed fully automated method consists of four parts; (i) individual tooth segmentation and identification module for IOS data (TSIM-IOS); (ii) individual tooth segmentation and identification module for CBCT data (TSIM-CBCT); (iii) global-to-local tooth registration between IOS and CBCT; and (iv) stitching error correction of full-arch IOS. The experimental results show that the proposed method achieved landmark and surface distance errors of 112.4$\mu$m and 301.7$\mu$m, respectively.
\end{abstract}


\section{Introduction}
\label{sec:introduction}
\IEEEPARstart{D}{ental} cone-beam computerized tomography (CBCT) and intraoral scan (IOS) have been used for implant treatment planning and maxillofacial surgery simulation. Dental CBCT has been widely used for the three-dimensional (3D) imaging of the teeth and jaws \cite{sukovic2003cone,miracle2009conebeam}. Recently, IOS has been increasingly used to capture digital impressions that are replicas of teeth, gingiva, palate, and soft tissue in the oral cavity \cite{mangano2017intraoral,zimmermann2015intraoral}, as digital scanning technologies have rapidly advanced \cite{robles2020digital}. The use of IOS addresses many of the shortcomings of the conventional impression  manufacturing techniques \cite{siqueira2021intraoral, manicone2021patient}.

This paper aims to provide a fully automated method of integrating dental CBCT and IOS data into one image such that the integrated image utilizes the strengths and supplements the weaknesses of each image. In dental CBCT, spatial resolution is insufficient for elaborately depicting tooth geometry and interocclusal relationships. Moreover, image degradation associated with metal-induced artifacts is becoming an increasingly frequent problem, as the number of older people with artificial dental prostheses and metallic implants is rapidly increasing with the rapidly aging populations. Metallic objects in the CBCT field of view produce streaking artifacts that highly degrade the reconstructed CBCT images, resulting in a loss of information on the teeth and other anatomical structures\cite{schulze2011artefacts}. IOS can compensate for the aforementioned weaknesses of dental CBCT. IOS provides 3D tooth crown and gingiva surfaces with a high resolution. However, tooth roots are not observed in intraoral digital impressions. Therefore, CBCT and IOS can be complementary to each other. A suitable fusion between CBCT and IOS images allows to provide detailed 3D tooth geometry along with the gingival surface.

\begin{figure*}	
	\centering
	\includegraphics[width=1\textwidth]{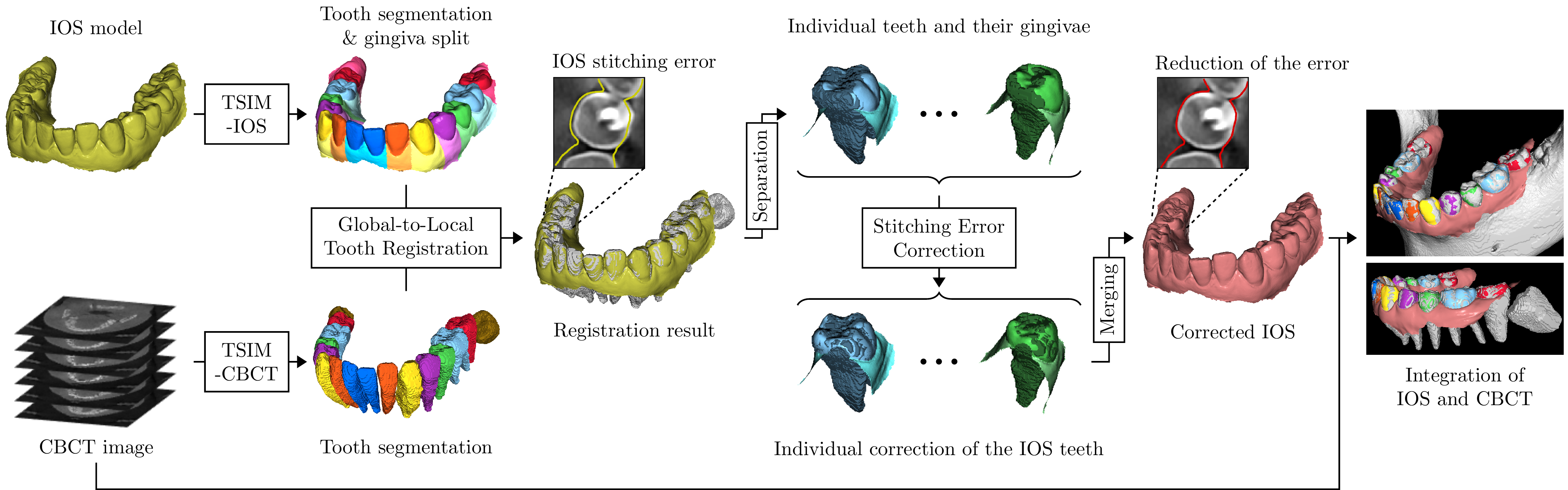}
	\caption{Overall flow diagram of the proposed method consisting of four parts; tooth segmentation and identification from IOS and CBCT data, global-to-local tooth registration of IOS and CBCT, and stitching error correction in IOS. Therefore, the proposed method integrates IOS and CBCT images into one coordinate system while improving the accuracy of full-arch IOS.}
	\label{fig:framework}
\end{figure*}

Numerous attempts have been made to register dental impression data to maxillofacial models obtained from 3D CBCT images. The registration process is to find a rigid transformation by taking advantage of the properties that the upper and lower jaw bones are rigid and the tooth surfaces are partially overlapping areas (\textit{e.g.}, the crowns of the exposed teeth). Several methods \cite{gateno2003new, uechi2006novel, swennen2007use, xia2009new, swennen2009cone} utilized fiducial markers for registration, which require a complicated process that involves the fabrication of devices with the markers, double CT scanning and post-processing for marker removal. To simplify these processes, virtual reference point-based methods \cite{kim2010integration, lin2013artifact, hernandez2013new, nilsson2016virtual} were proposed to roughly align two models using reference points, and achieve a precise fit by employing an iterative closest point (ICP) method \cite{besl1992method}. ICP is a widely used iterative registration method consisting of the closest point matching between two data and minimization of distances between the paired points. However, the ICP method relies heavily on initialization because it can easily be trapped into a local optimal solution. Therefore, these methods based on ICP typically require the user-involved initial alignment, which is a cumbersome and time-consuming procedure due to manual clicking. Furthermore, registration using ICP can be difficult to achieve acceptable results for patients containing metallic objects \cite{flugge2017registration}. Teeth in CBCT images that are contaminated by metal artifacts prevent accurate point matching with teeth in impressions. Therefore, there is a high demand for a fully automated and robust registration method. Recently, a deep learning-based method \cite{chung2020automatic} was used to automate the initial alignment by extracting pose cues from two data. This approach has limitations in achieving sufficient registration accuracy enough for clinical application. Without the use of a very good initial guess, the point matching for multimodal image registration is affected by the non-overlapping area of the two different modality data (\textit {e.g.}, the soft tissues in IOS and the jaw bones and tooth surfaces contaminated by metal artifacts in CBCT).

For an accurate registration, it is necessary to separate the non-overlapping areas as much as possible to prevent incorrect point matching. Therefore, individual tooth segmentation and identification in CBCT and IOS are required as important preprocessing tasks. In recent years, owing to advances in deep learning methods, numerous fully automated 3D tooth segmentation methods have been developed for CBCT images \cite{lee2020automated,rao2020symmetric,chen2020automatic,cui2019toothNet,jang2021fully} and impression models \cite{lian2020deep,zanjani2021mask,cui2021tsegnet}.

Although the performance of intraoral scanners is improving, full-arch scans have not yet surpassed the accuracy of conventional impressions \cite{zhang2021accuracy,giachetti2020accuracy}. IOS at short distances is available to obtain partial digital impressions that can replace traditional dental models, but it may not yet be suitable for clinical use on long complete-arches due to the global cumulative error introduced during the local image stitching process \cite{ender2019accuracy}. To achieve sophisticated image fusion, it is therefore necessary to correct the stitching errors of IOS.

We propose a fully automated method for registration of CBCT and IOS data as well as correction of IOS stitching errors. The proposed method consists of four parts: (i) individual tooth segmentation and identification module from IOS data (TSIM-IOS); (ii) individual tooth segmentation and identification module from CBCT data (TSIM-CBCT); (iii) global-to-local tooth registration between IOS and CBCT; and (iv) stitching error correction of the full-arch IOS. We developed TSIM-IOS using 2D tooth feature-highlighted images, which are generated by orthographic projection of the IOS data. This approach allows high-dimensional 3D surface models to efficiently segment individual teeth using low-dimensional 2D images. In TSIM-CBCT, we utilize the panoramic image-based deep learning method \cite{jang2021fully}. This method is robust against metal artifacts because it utilizes panoramic images (generated from CBCT images) not significantly affected by metal artifacts. The TSIM-IOS and -CBCT are used to focus only on the teeth, while removing as many non-overlapping areas as possible. In (iii), we then align the two highly overlapping data (\textit{i.e.}, the segmented teeth in the CBCT and IOS data) through global-to-local fashion, which consists of global initialization by fast point feature histograms (FPFH) \cite{rusu2009fast} and local refinement by ICP based on individual teeth (T-ICP). T-ICP allows the closest point matching only between the same individual teeth in the CBCT and IOS. The last part (iv) corrects the stitching errors of IOS using CBCT-derived tooth surfaces. Owing to the reliability of CBCT \cite{baumgaertel2009reliability}, location information of 3D teeth in the CBCT data can be used as a reference for correction of IOS teeth. After registration, each IOS tooth is fixed through a slightly rigid transformation determined by the reference CBCT tooth.

The main contributions of this paper are summarized as follows.

\begin{itemize}
	\item To the best of our knowledge, this study is the first to provide a sophisticated fusion of IOS and CBCT data at the level of accuracy required for clinical use.
	\item The proposed method can provide accurate intraoral digital impressions that correct cumulative stitching errors.
	\item This framework is robust against metal-induced artifacts in low-dose dental CBCT.
	\item The combined tooth-gingiva models with individually segmented teeth can be used for occlusal analysis and surgical guide production in digital dentistry.
\end{itemize}

The remainder of this paper is organized as follows. Section 2 describes the proposed method in detail. In Section 3, we explain the experimental results. Section 4 presents the discussion and conclusions.

\section{Method}

The overall framework of the proposed method is illustrated in Fig. \ref{fig:framework}. It is designed to automatically align a patient's IOS model with the same patient's CBCT image. IOS models consist of 3D surfaces (triangular meshes) of the upper and lower teeth, and are acquired in Standard Triangle Language (STL) file format containing 3D coordinates of the triangle vertices. The 3D vertices of IOS data can be expressed as a set of 3D points and the measurement unit of these points is millimeter. Dental CBCT images are isotropic voxel structures consisting of sequences of 2D cross-sectional images, and are saved in Digital Imaging and Communications in Medicine (DICOM) format. 

Registration between two different imaging protocols must be separately obtained for the maxilla and mandible. For convenience, only the method for the mandible is described in this section. The method for the maxilla is the same.

\subsection{Individual Tooth Segmentation and Identification in IOS} \label{subsec:IOS}

\begin{figure}[ht]
	\centering
	\subfloat[]{\includegraphics[width=.2\textwidth]{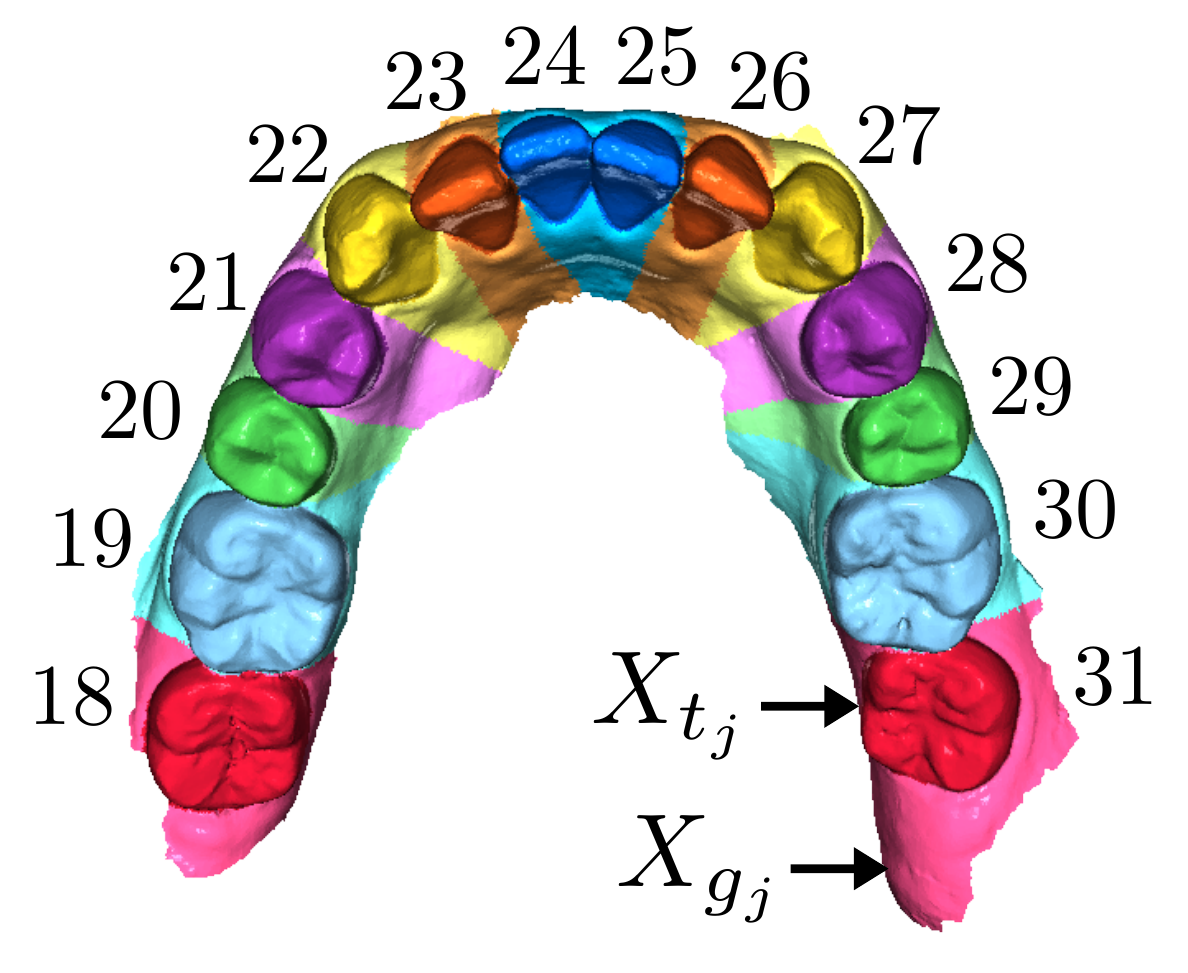}}~~~
	\subfloat[]{\includegraphics[width=.225\textwidth]{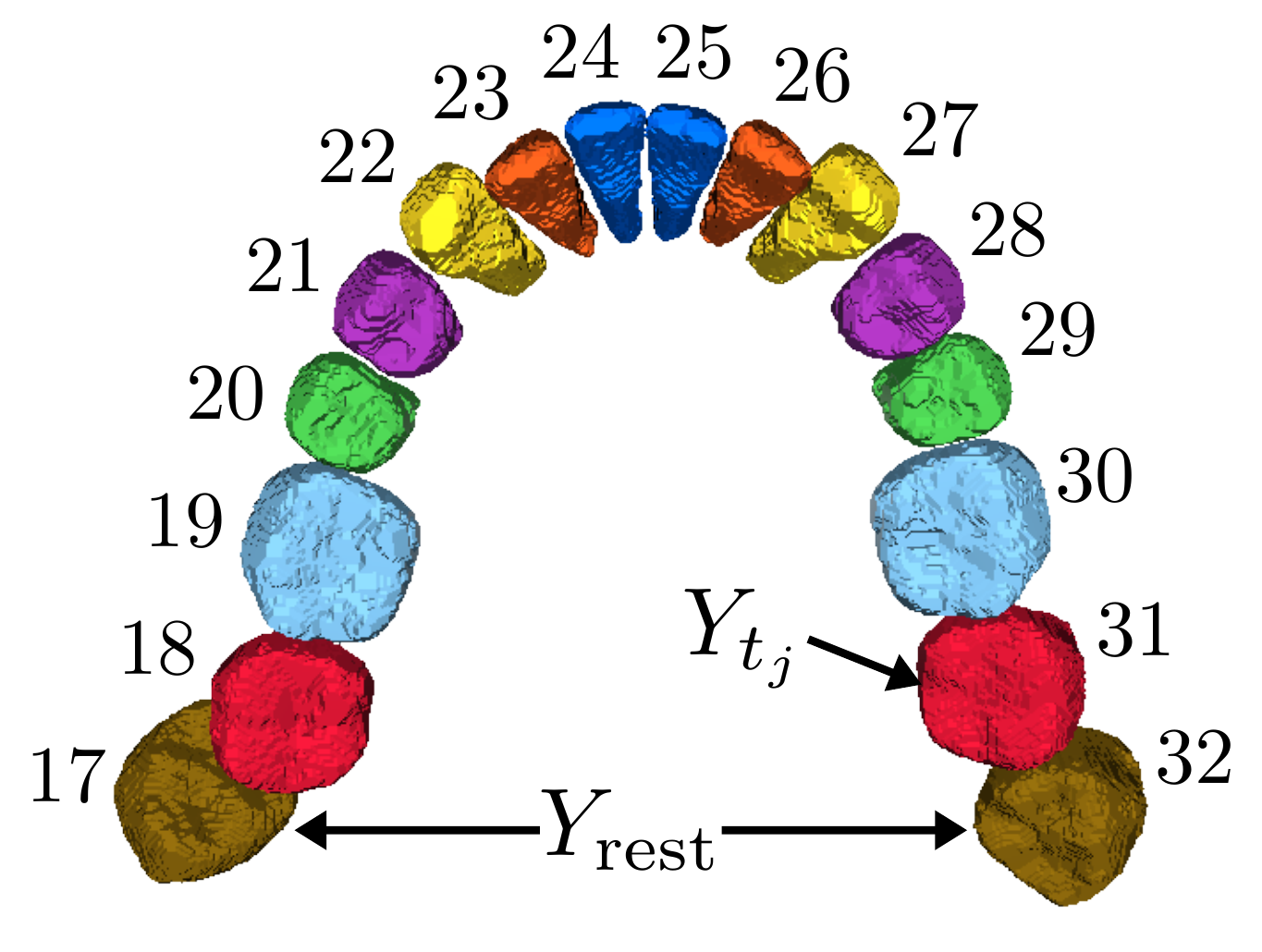}}
	\caption{Results of TSIM-IOS and -CBCT, respectively. The indicated numbers represent mandibular teeth by the universal notation. (a) Individual IOS teeth and their split gingiva parts, and (b) CBCT teeth containing unexposed wisdom teeth.}
	\label{fig:segid}
\end{figure}

As shown in Fig. \ref{fig:segid}a, TSIM-IOS decomposes the 3D point set $X$ of the IOS model into

\begin{equation}
	X = \underbrace{X_{t_1} \cup \cdots \cup X_{t_J}}_{X_{\mbox{\scriptsize teeth}}}\cup X_{\mbox{\scriptsize gingiva}},
\end{equation}
where each $X_{t_j}$ represents a tooth with the code $t_j$ in $X$, $J$ is the number of teeth in $X$, and $X_{\mbox{\scriptsize gingiva}}$ is the rest including the gingiva in $X$. According to the universal notation system \cite{nelson2014wheeler}, $t_j$ is the number between 1 and 32 that is assigned to an individual tooth to identify the unique tooth. A detailed explanation is provided in \ref{app:sec1}.

Additionally, we divide the gingiva $X_{\mbox{\scriptsize gingiva}}$ into 

\begin{equation} \label{eq:gingiva}
	X_{\mbox{\scriptsize gingiva}} = X_{g_1} \cup \cdots \cup X_{g_J},
\end{equation}
where 

\begin{equation}
	X_{g_j} = \left\{\x \in X_{\mbox{\scriptsize gingiva}} : \underset{\x' \in X_{\mbox{\tiny teeth}}}{\mbox{argmin}}\| \x-\x'\| \in X_{t_j}\right\}	.
\end{equation}
Therefore, a point in $X_{\mbox{\scriptsize gingiva}}$ belongs to a separated gingiva $X_{g_j}$ according to the nearest tooth $X_{t_j}$.

\subsection{Individual Tooth Segmentation and Identification in CBCT} \label{subsec:CBCT}

TSIM-CBCT is based on a deep learning-based individual tooth segmentation and identification method developed by Jang \textit{et al.} \cite{jang2021fully}. As shown in Fig. \ref{fig:segid}b, we obtain the teeth point cloud $Y$ that consists of individual tooth point clouds, denoted by

\begin{equation}
	Y = \underbrace{Y_{t_1} \cup \cdots \cup Y_{t_J}}_{Y_{\mbox{\scriptsize teeth}}} \cup Y_{\mbox{\scriptsize rest}},
\end{equation}
where each $Y_{t_j}$ represents the $t_j$-tooth for $j=1,\cdots,J$ and $Y_{\mbox{\scriptsize rest}}$ refers to a point cloud of unexposed teeth (\textit{e.g.}, impacted wisdom teeth) if presented. Because the impacted teeth do not appear in IOS images, they are separated by $Y_{\mbox{\scriptsize rest}}$.

Each tooth point cloud $Y_{t_j}$ is obtained from a 3D binary image of the $t_j$-tooth determined by the individual tooth segmentation and identification method \cite{jang2021fully}. The points in $Y_{t_j}$ lie on isosurfaces (approximating the boundary of the segmented tooth image) that are generated by the marching cube algorithm \cite{lewiner2003efficient}. Because the unit of points in $Y_{t_j}$ is associated with the image voxels, the points are scaled in millimeters by the image spacing and slice thickness.

\subsection{Global-to-Local Tooth Registration of IOS and CBCT} 
This subsection describes the registration method to find the optimal transformation $\mathcal{T}^*$ such that the transformed point cloud $\mathcal{T}^*(X) = \{\mathcal{T}^*(\x) : \x \in X \}$ best aligns with the target $Y$ in terms of partially overlapping tooth surfaces. The registration problem consists of the following two steps:
\begin{enumerate}
\item Construct a set of correspondences $Corr= \{(\x,\y)\in X \times Y\}$ between a source $X$ and target $Y$.
\item Find the optimal rigid transformation with the following mean square error minimization to best match the pairs in the correspondences

	\begin{equation}
		\mathcal{T}^* = \underset{\mathcal{T} \in SE(3)}{\mbox{argmin}} \sum_{(\x,\y) \in Corr} \|\y-\mathcal{T}(\x)\|^2,
	\end{equation}

	where $SE(3)$ is the set of rigid transformations that are modeled with a $4\times4$ matrix determined by three angles and a translation vector.
\end{enumerate}
Here, we adopt two registration methods: FPFH \cite{rusu2009fast} for global initial alignment, and an improved ICP using individual tooth segmentation for local refinement.

\subsubsection{Global initial alignment of the IOS and CBCT teeth}
We compute the two sets of FPFH vectors \cite{rusu2009fast}; $\mbox{FPFH}(X_{\mbox{\scriptsize teeth}})=\{\mbox{FPFH}(\x):\x \in X_{\mbox{\scriptsize teeth}}\}$ and $\mbox{FPFH}(Y_{\mbox{\scriptsize teeth}})=\{\mbox{FPFH}(\y):\y \in Y_{\mbox{\scriptsize teeth}}\}$. $\mbox{FPFH}(\x)$ represents not only the geometric features of the normal vector and the curvature at $\x \in X_{\mbox{\scriptsize teeth}}$, but also the relevant information considering its neighboring points over $X_{\mbox{\scriptsize teeth}}$. The details of FPFH are provided in \ref{app:sec2}. 

$\mbox{FPFH}(X_{\mbox{\scriptsize teeth}})$ and $\mbox{FPFH}(Y_{\mbox{\scriptsize teeth}})$ are used to find correspondences between $X_{\mbox{\scriptsize teeth}}$ and $Y_{\mbox{\scriptsize teeth}}$. For each $\x \in X_{\mbox{\scriptsize teeth}}$, we select $\y \in Y_{\mbox{\scriptsize teeth}}$, denoted by $\mbox{match}^{\mbox{\tiny FPFH}}_{Y_{\mbox{\tiny teeth}}}(\x)$, whose FPFH vector is most similar to $\mbox{FPFH}(\x)$:

\begin{equation}
	\mbox{match}^{\mbox{\tiny FPFH}}_{Y_{\mbox{\tiny teeth}}}(\x)=\underset{\y \in Y_{\mbox{\tiny teeth}}}{\text{argmin}}~{\|\mbox{FPFH}(\x)-\mbox{FPFH}(\y)\|}.
\end{equation}
Similarly, we compute $\mbox{match}^{\mbox{\tiny FPFH}}_{X_{\mbox{\tiny teeth}}}(\y)$ for all $\x \in X_{\mbox{\scriptsize teeth}}$. Then, we obtain the correspondence set

\begin{equation}
	Corr = Corr_{X_{\mbox{\tiny teeth}}} \cap Corr_{Y_{\mbox{\tiny teeth}}},
\end{equation}
where

\begin{align}
	&Corr_{X_{\mbox{\tiny teeth}}} = \left\{\left(\x,\mbox{match}^{\mbox{\tiny FPFH}}_{Y_{\mbox{\tiny teeth}}}(\x)\right):\x \in X_{\mbox{\scriptsize teeth}} \right\},\\ 
	&Corr_{Y_{\mbox{\tiny teeth}}} = \left\{\left(\mbox{match}^{\mbox{\tiny FPFH}}_{X_{\mbox{\tiny teeth}}}(\y),\y \right):\y \in Y_{\mbox{\scriptsize teeth}} \right\}.
\end{align}
The set $Corr$ contains pairs $(\x,\y) \in X_{\mbox{\scriptsize teeth}} \times Y_{\mbox{\scriptsize teeth}}$ where $\mbox{FPFH}(\x)$ and $\mbox{FPFH}(\y)$ are the most similar to each other. However, such simple feature information alone cannot provide a proper point matching between $X_{\mbox{\scriptsize teeth}}$ and $Y_{\mbox{\scriptsize teeth}}$, because there are too many points with similar geometric features in the point clouds. To filter out inaccurate pairs from the set ${Corr}$, we randomly sample three pairs $(\x_1,\y_1)$, $(\x_2,\y_2)$, $(\x_3,\y_3)\in {Corr}$ and select them if the following conditions \cite{zhou2016fast} are met, and drop them otherwise:

\begin{equation} \label{eq:filter}
	\tau < \frac{\|\x_i - \x_j\|}{\|\y_i - \y_j\|} < \frac{1}{\tau},~~\text{for}~1\leq i<j \leq 3,
\end{equation}
where $\tau$ is a number close to 1. We denote this filtered subset as ${Corr}^{(0)}$. Then, the initial transformation is determined by

\begin{equation}
	\mathcal{T}^{(0)}=\underset{\mathcal{T} \in SE(3)}{\mbox{argmin}} \sum_{(\x,\y) \in Corr^{(0)}} \|\y-\mathcal{T}(\x)\|^2.
\end{equation}

\subsubsection{Local refinement of the roughly aligned teeth}

We denote $X_{\mbox{\scriptsize teeth}}$ transformed by the previously obtained $\mathcal{T}^{(0)}$ as $X_{\mbox{\scriptsize teeth}}^{(0)} = X_{t_1}^{(0)} \cup \cdots \cup X_{t_J}^{(0)}$, where $X_{t_j}^{(0)} =  \mathcal{T}^{(0)}(X_{t_j})$ for $j=1,\cdots,J$. $X_{\mbox{\scriptsize teeth}}^{(0)}$ and $Y_{\mbox{\scriptsize teeth}}$ are then roughly aligned, but fine-tuning is needed to achieve accurate registration. A fine rigid transformation is obtained through an iterative process, which gradually improves the correspondence finding. We propose an improved ICP (T-ICP) method with point matching based on individual teeth.

For $k \geq 1$, we denote $X_{\mbox{\scriptsize teeth}}^{(k)} = \mathcal{T}^{(k)}(X_{\mbox{\scriptsize teeth}}^{(k-1)})$. Here, the $k$-th rigid transformation $\mathcal{T}^{(k)}$ is determined by

\begin{equation}
	\mathcal{T}^{(k)} = \underset{\mathcal{T} \in SE(3)}{\mbox{argmin}} \sum_{(\x,\y) \in Corr^{(k)}} \|\y-\mathcal{T}(\x)\|^2.
\end{equation}
The correspondence set $Corr^{(k)}$ for $k$ is given by

\begin{equation}
	Corr^{(k)} = \left\{ \left(\x, \mbox{match}_{Y_{\mbox{\tiny teeth}}}(\x) \right) : \x \in X_{\mbox{\scriptsize teeth}}^{(k-1)} \right\} \cap P^{(k)},
\end{equation}
where

\begin{align}
	& \mbox{match}_{Y_{\mbox{\tiny teeth}}}(\x)=\underset{\y \in Y_{\mbox{\tiny teeth}}}{\mbox{argmin}} \|\x-\y\|, \\
	& P^{(k)} = \bigcup_{j=1}^n \left\{(\x,\y) \in X_{t_j}^{(k-1)} \times Y_{t_j} \right\}.
\end{align}
Using the set $P^{(k)}$ prevents undesired correspondences between two teeth with different codes. Note that this is the vanilla ICP when $P^{(k)}$ is not used. The final rigid transformation $\mathcal{T}^*$ is obtained by the following composition of transformations: $\mathcal{T}^*=\mathcal{T}^{(K)} \circ \cdots \circ \mathcal{T}^{(0)}$, where $K$ is the number of iterations until the stopping criterion is satisfied for a given $\varepsilon>0$:

\begin{equation}
	\sum_{(\x,\y) \in Corr^{(K)}} \| \mathcal{T}^{(K)} \circ \cdots \circ \mathcal{T}^{(0)}(\x) - \y \|<\varepsilon.
\end{equation}

\subsection{Stitching Error Correction in IOS}

Next, we edit the IOS models with stitching errors by referring to the CBCT images. We denote $X_{t_j}^*=\mathcal{T}^*(X_{t_j})$ and $X_{g_j}^*=\mathcal{T}^*(X_{g_j})$ for $j=1,\cdots,J$. Each tooth $X^*_{t_j}$ is transformed by a corrective rigid transformation $\mathcal{T}_j^{**}$, which is obtained by applying the vanilla ICP to sets $X^*_{t_j-1} \cup X^*_{t_j} \cup X^*_{t_j+1}$ and $Y^*_{t_j-1} \cup Y^*_{t_j} \cup Y^*_{t_j+1}$ as the source and target. Here, $X^*_{t_j-1}$ (or $X^*_{t_j+1}$) is an empty set if $t_j-1$ (or $t_j+1$) is not equal to $t_{j'}$ for every $j'=1,\cdots,J$. Using the individual corrective transformations, IOS stitching errors are corrected separately by $X_{t_j}^{**}=\mathcal{T}_{j}^{**}(X_{t_j}^{*})$ for $j=1,\cdots,J$. In this procedure, we use one tooth and two adjacent teeth on both sides for reliable correction. It takes advantage of the fact that narrow digital scanning is accurate. Now it remains to fix the gingiva area whose boundary shares the boundaries with the teeth. To fit the boundaries between the gingiva and individually transformed teeth, the gingival surface is divided according to the areas in contact with the individual teeth by Eq. \eqref{eq:gingiva}. Therefore, the rectified gingiva is obtained by $X_{g_j}^{**} = \mathcal{T}_{j}^{**}(X_{g_j}^{*})$ for $j=1,\cdots,J$.

\section{Experiments and Results}

Experiments were carried out using CBCT images in DICOM format and IOS models in STL format. Each CBCT image is produced by a dental CBCT machine: DENTRI-X (HDXWILL), which uses tube voltages of 90kVp and a tube current of 10mA. The size of images obtained by the machine is $800\times800\times400$. The pixel spacing and slice thickness are both $0.2$mm. Each IOS model is scanned by one of two intraoral scanners: i500 (Medit) and TRIOS 3 (3shape). An IOS model is either maxilla or mandible, which has approximately 120,000 vertices and 200,000 triangular faces, respectively. The dataset were provided by HDXWILL. Additionally, we used maxillary and mandibular digital dental models to train TSIM-IOS. These dataset were collected by the Yonsei University College of Dentistry. Personal information in all dataset was de-identified for patient privacy and confidentiality.

\begin{table*}[h]
	\footnotesize
	\centering
	\caption{Identification Results of TSIM-IOS and CBCT.}
	\label{tbl:eval_tsim_id}
	\vskip 0.0in
	\begin{tabular}{cccccccccc} \hline
		{} & \multicolumn{3}{c}{\bf TSIM-IOS} & \multicolumn{3}{c}{\bf TSIM-CBCT} \\
		{} & {\bf Precision (\%)} & {\bf Recall (\%)} & {\bf F1-score (\%)} & {\bf Precision (\%)} & {\bf Recall (\%)} & {\bf F1-score (\%)}\\ 
		\hline
		{Central Incisors} & {97.11} & {95.29} & {96.19} & {96.29} & {92.91} & {94.58} \\
		{Lateral Incisors} & {97.24} & {93.11} & {95.14} & {94.33} & {95.23} & {94.77} \\
		{Canines} & {95.15} & {93.34} & {94.24} & {95.58} & {98.12} & {96.84} \\
		{1st Premolars} & {97.03} & {89.91} & {93.34} & {96.44} & {90.2} & {93.21} \\
		{2nd Premolars} & {95.05} & {94.95} & {95.01} & {95.74} & {92.68} & {94.19} \\
		{1st Molars} & {94.04} & {91.31} & {92.67} & {96.50} & {91.86} & {94.12} \\
		{2nd Molars} & {96.96} & {90.43} & {93.58} & {95.67} & {94.08} & {94.87}\\
		{3rd Molars} & {$-$} & {$-$} & {$-$} & {97.25} & {93.92} & {95.55}\\						
		\hline
		{\bf Mean} & {\bf 96.08 $\pm$ 1.20} & {\bf 93.62 $\pm$ 1.97} & {\bf 95.31 $\pm$ 1.13} & {\bf 95.97 $\pm$ 0.81} & {\bf 93.63 $\pm$ 2.22} & {\bf 94.76 $\pm$ 1.02}\\
		\hline
	\end{tabular}
\end{table*}

\begin{figure*}
	\centering
	\subfloat[]{\includegraphics[width=.18\textwidth]{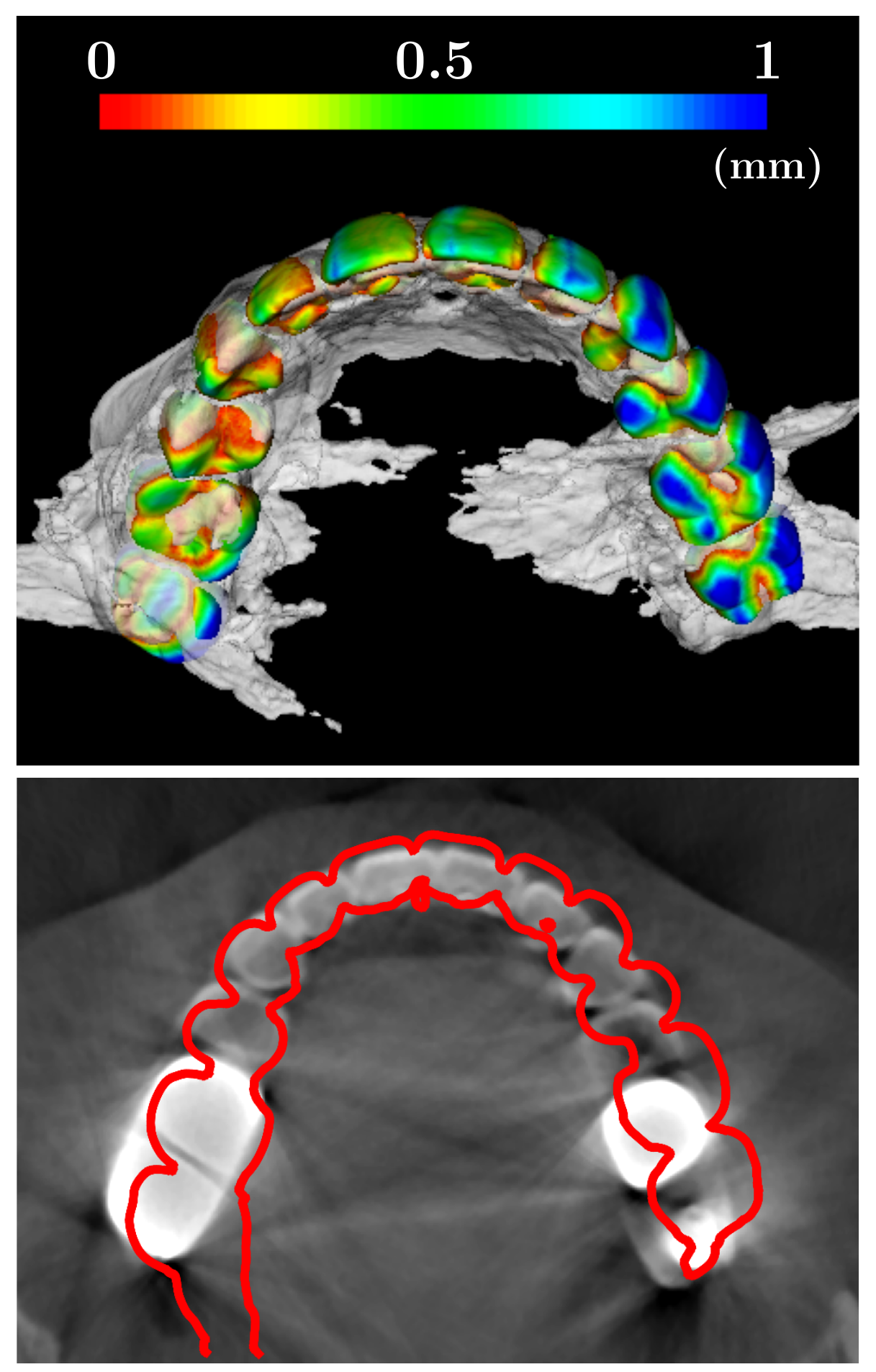}}~
	\subfloat[]{\includegraphics[width=.18\textwidth]{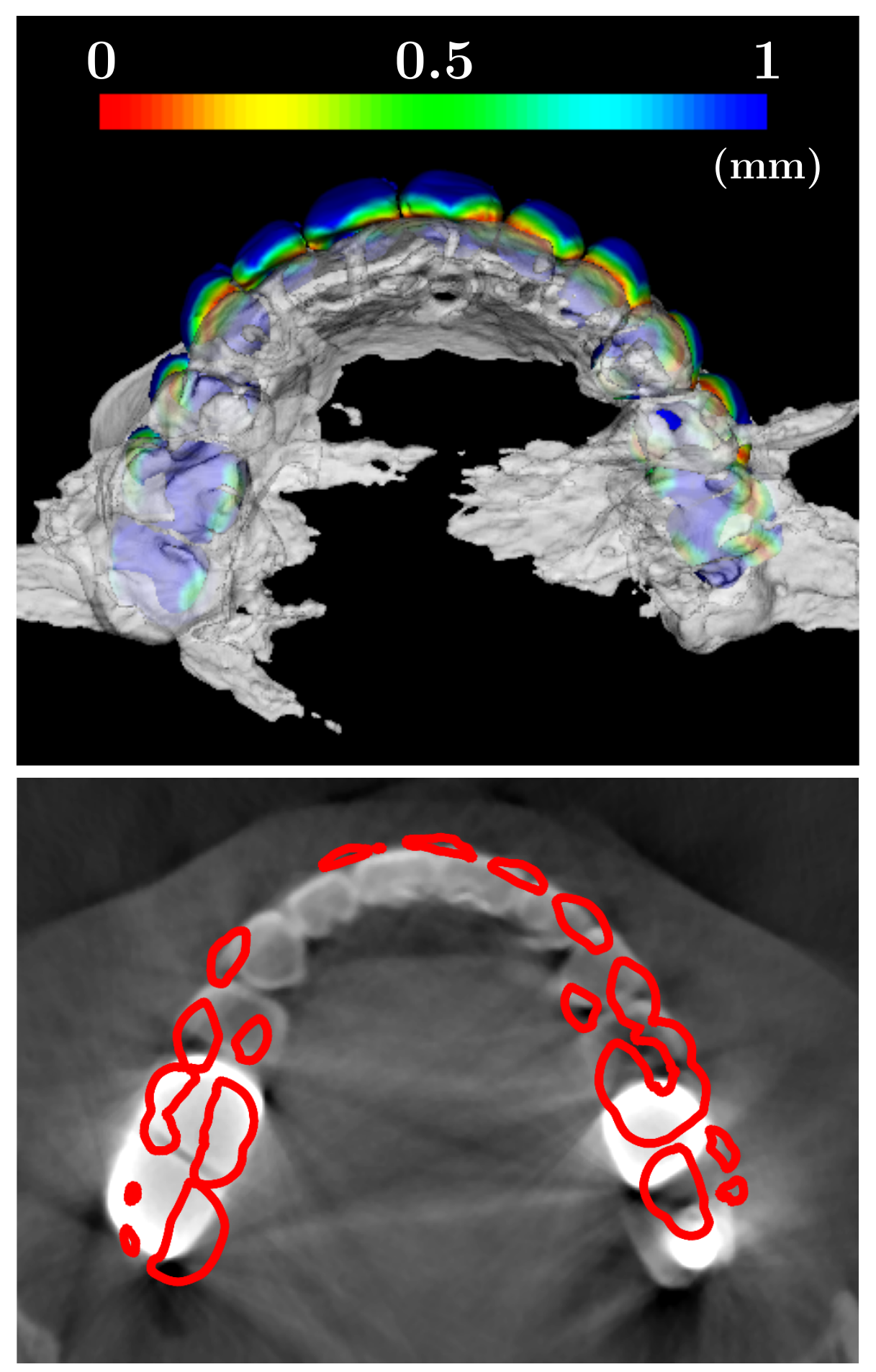}}~
	\subfloat[]{\includegraphics[width=.18\textwidth]{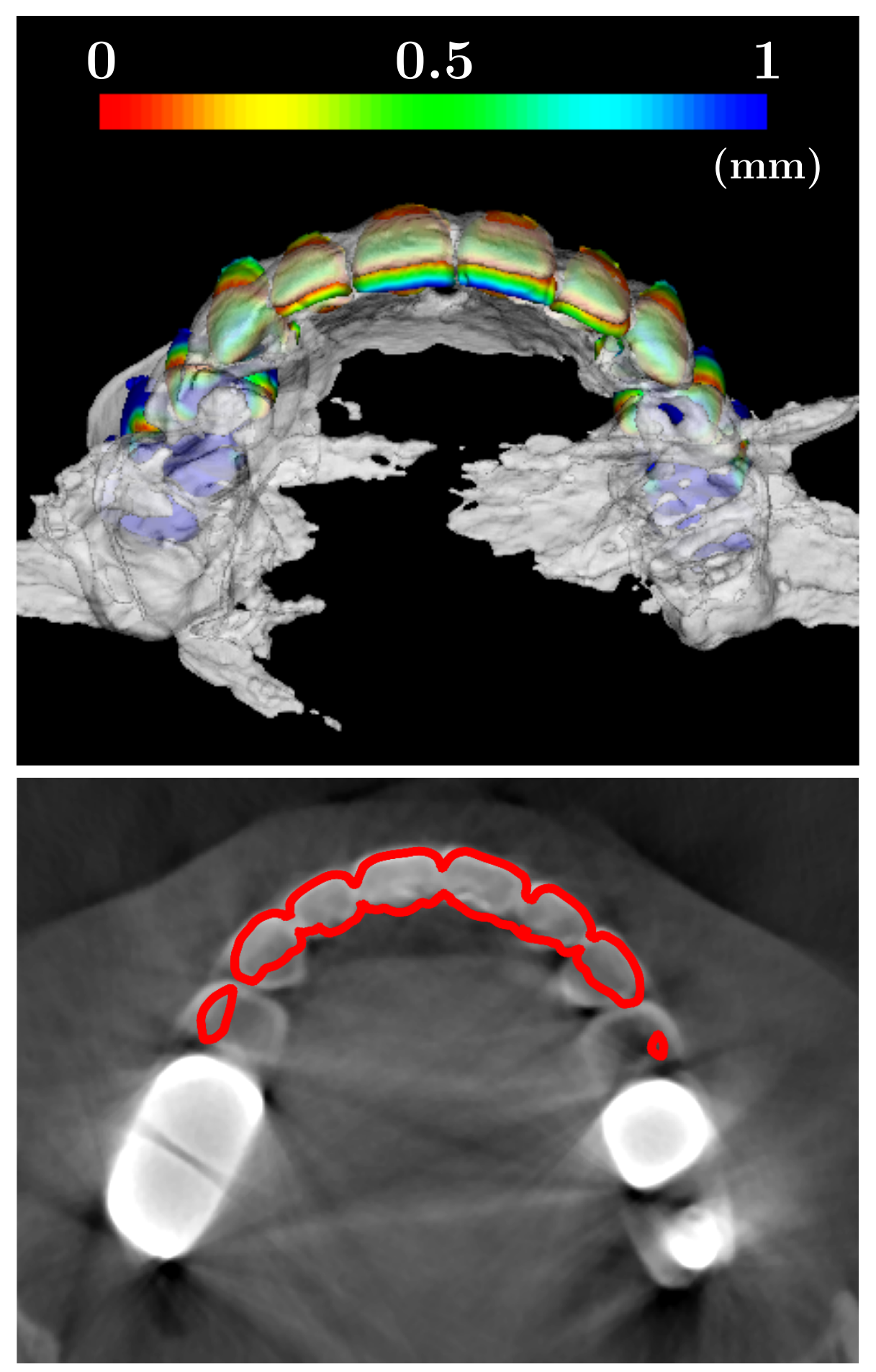}}~
	\subfloat[]{\includegraphics[width=.18\textwidth]{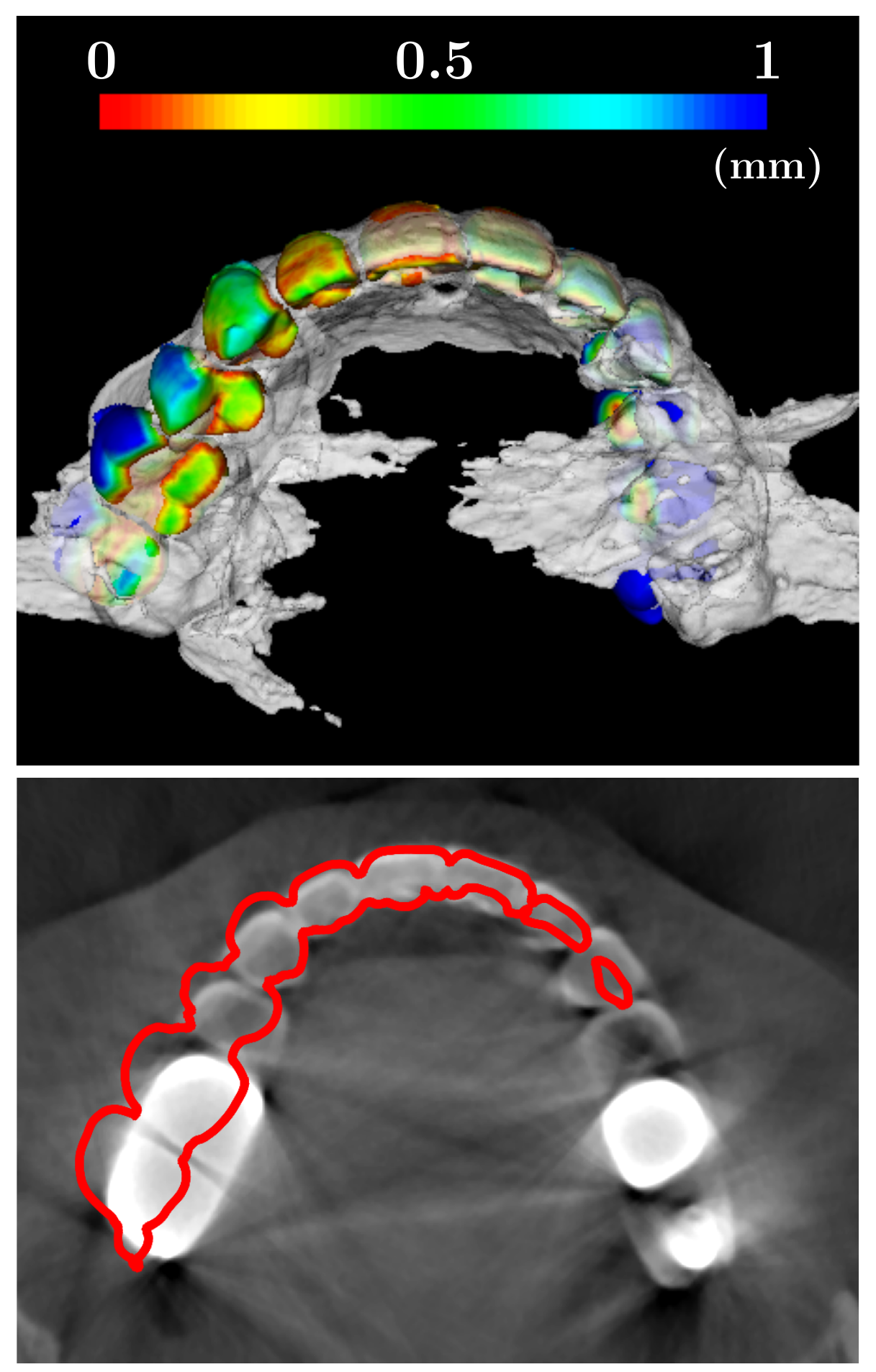}}~
	\subfloat[]{\includegraphics[width=.18\textwidth]{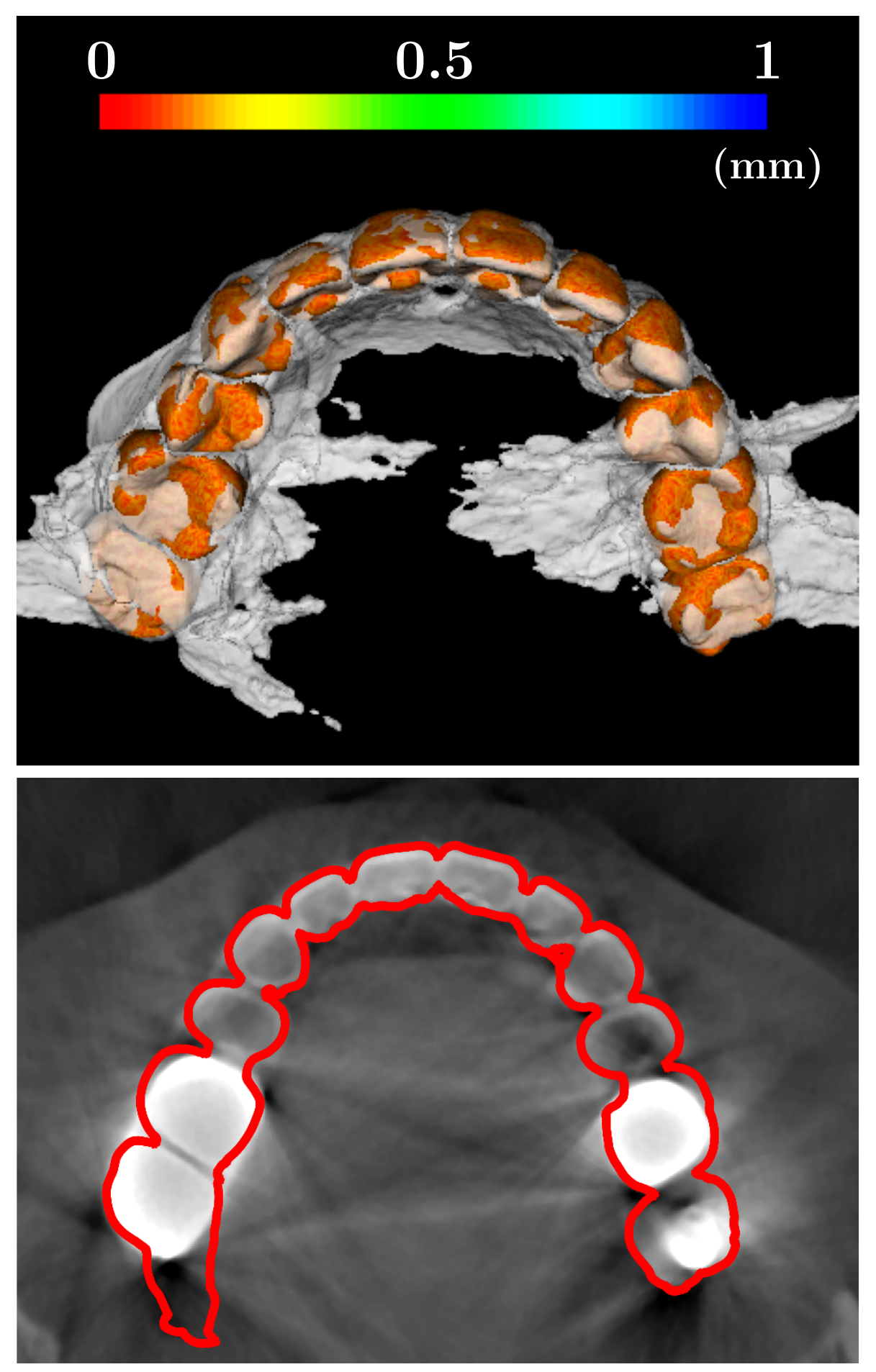}}
	
	\caption{Qualitative comparison results of registration methods. (a) MR, (b) CPD, (c) FPFH, (d) FPFH followed by ICP, and (e) the proposed method. In the 3D visualization of the results, the colors on the IOS teeth represent distances between the IOS and CBCT tooth surfaces. On the 2D CT slice images, the red contours are cross sections of the aligned IOS models cut along the corresponding CT slice.}
	\label{fig:quantitative_reg_result}
\end{figure*}

\begin{table}[]
	\footnotesize
	\centering
	\caption{Segmentation Results of TSIM-IOS and CBCT.}
	\label{tbl:eval_tsim_seg}
	\vskip 0.0in
	\begin{tabular}{ccccccc} \hline
		\multicolumn{2}{c}{\bf TSIM-IOS} & \multicolumn{2}{c}{\bf TSIM-CBCT} \\
		{\bf DSC (\%)} & {\bf Accuracy (\%)} & {\bf DSC (\%)} & {\bf Accuracy (\%)} \\ 
		\hline
		{$95.10 \pm 1.59$}  & {$95.45 \pm 1.32$} & {$95.28 \pm 1.51$}  & {$98.35 \pm 0.34$}\\
		\hline
	\end{tabular}
\end{table}

\subsection{Results of TSIM-IOS and CBCT}

In this subsection, we present the results of the deep learning-based individual tooth segmentation and identification methods: TSIM-IOS and CBCT in Sections \ref{subsec:IOS} and \ref{subsec:CBCT}, respectively. For TSIM-IOS, 71 maxillary and mandibular dental models were used for training and 35 models for testing. Similarly, in TSIM-CBCT, 49 3D CBCT images were used for training and 23 images for testing.

To evaluate the tooth identification performance, we used three metrics: precision, recall, and F1-score. Table \ref{tbl:eval_tsim_id} provides the quantitative evaluation for TSIM-IOS and CBCT. In the TSIM-IOS, we intentionally excluded the third molar (\textit{i.e.}, wisdom teeth) as their surfaces may not be entirely exposed beyond gingiva. Similarly, Table \ref{tbl:eval_tsim_seg} summarizes the result of individual tooth segmentation using two metrics: Dice similarity coefficient (DSC) and accuracy.

\subsection{Evaluation and Result of the Proposed Registration Method}
We used 22 pairs of IOS models and CBCT images to evaluate the performance of the proposed registration method. Each pair was obtained from the same patient. Sixteen of the subjects had one or more metallic objects such as dental restoration, while the remaining six patients were metal-free.

To validate the registration accuracy, we used landmark and surface distances between IOS and CBCT data. The landmark distance is the mean of distances between corresponding points in IOS and CBCT data:

\begin{equation}\label{eq:land}
	E_{land}(\hat{X},\hat{Y};\mathcal{T}) = \frac{1}{N}\sum_{i=1}^N \| \mathcal{T}(\hat{\x}_i)-\hat{\y}_i\|, 
\end{equation}
where $\mathcal{T}$ is a rigid transformation and, $\hat{X}=\{\hat{\x}_1,\cdots,\hat{\x}_N\}$ and $\hat{Y}=\{\hat{\y}_1,\cdots,\hat{\y}_N\}$ are the landmark sets of the pair of IOS and CBCT data, respectively. The surface distance is from the IOS tooth surfaces to the CBCT tooth surfaces:

\begin{equation}\label{eq:surf}
	E_{surf}(\bar{X},\bar{Y};\mathcal{T}) = \sup_{\bar{\x} \in \bar{X}} \inf_{\bar{y} \in \bar{Y}} \| \mathcal{T}(\bar{\x})-\bar{\y} \|, 
\end{equation}
where $\bar{X}$ and $\bar{Y}$ are the tooth surfaces of IOS and CBCT data, respectively. To make the ground truths, tooth landmark annotation and segmentation labeling were carefully conducted by dentists.

To show the effectiveness of the proposed method, we compared it manual clicking registration with ICP (MR), coherent point drift (CPD) \cite{myronenko2010point}, FPFH, and FPFH followed by ICP. These methods are implemented using raw IOS model and skull model, which is obtained by applying thresholding segmentation and the marching cube algorithm to CBCT images. Table \ref{tbl:eval_reg} provides the quantitative evaluations of the methods, and Fig. \ref{fig:quantitative_reg_result} displays the qualitative results by visualizing distance maps between the ground-truth tooth surfaces of CBCT and IOS, which are aligned by rigid transformations obtained from the employed methods. Also, we performed an ablation study to present the advantage of TSIM-IOS and CBCT, as reported in Table \ref{tbl:eval_reg}.

\begin{figure}[h]
	\centering
	\includegraphics[width=.38\textwidth]{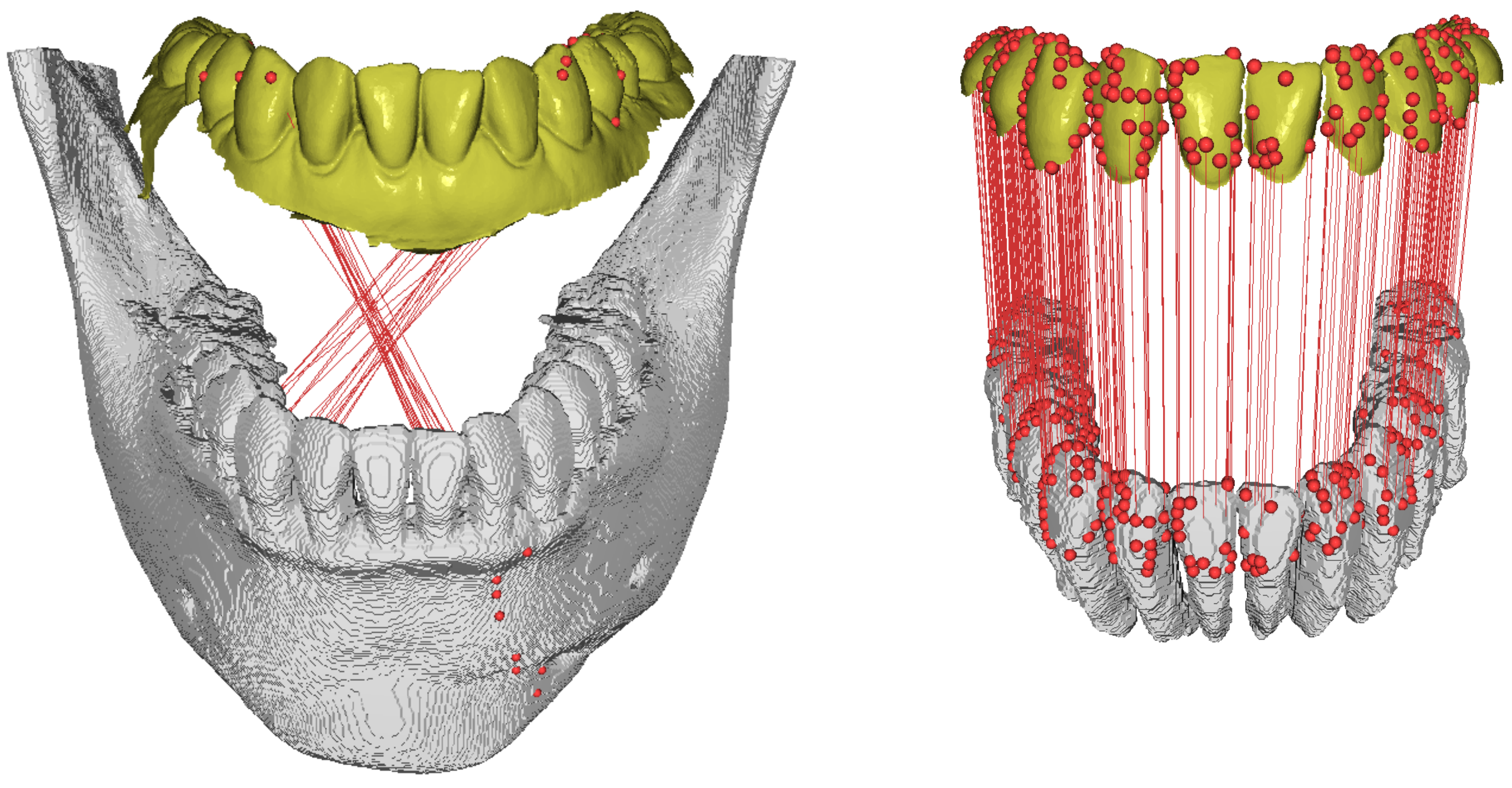}
	\caption{Correspondence pairs of FPFH-based methods. The image on the left shows poor matching from the FPFH method without TSIM. On the other hand, the image on the right shows modest correspondences between the teeth obtained by TSIM.}
	\label{fig:comparison_fpfh}
\end{figure}

When source and target point clouds partially overlap, the MR and CPD were less accurate than FPFH, suggesting that the feature-based method is more suitable than user interaction and probabilistic-based methods. But above all, these methods suffer from the unnecessary points because non-overlapping areas between the IOS and skull models (\textit{i.e.}, alveolar bones in CBCT and soft tissues in IOS) occupy the most of the entire area. In such condition, FPFH may produce inaccurate correspondence pairs due to the non-overlapping points that are not properly filtered out in Eq. \eqref{eq:filter}, as shown in Fig. \ref{fig:comparison_fpfh}. Therefore, the use of TSIM-IOS and -CBCT is beneficial by eliminating the areas that may adversely affect accurate registration. In the ablation study, the methods with TSIM showed improved performances compared to those without TSIM. Still, the MR and CPD have limitations in achieving automation and improving accuracy due to the roots of CBCT teeth, respectively. To precisely match the models roughly aligned by FPFH, we developed T-ICP, which is an improved ICP method that uses individual tooth segmentation. Adopting T-ICP instead of ICP led to increased accuracy. The advantage of T-ICP is that it avoids point correspondences between adjacent teeth with different codes. This constraint prevents unwanted correspondences by performing point matching only between the same CBCT and IOS teeth.

\begin{table}[]
	\footnotesize
	\centering
	\caption{Quantitative Comparison Results of Registration Methods. \label{tbl:eval_reg}}\vskip 0.05in
	\begin{tabular}{ccccccc} \hline
		{\bf Method} &{\textbf{Landmark} ($\mu$m)} & {\textbf{Surface} ($\mu$m)} \\ \cline{1-3}
		{MR} & {$1472.6 \pm 2401.7$}  & {$ 3117.4 \pm 3676.8$}\\
		{CPD} & {$12773.7 \pm 6123.6$}  & {$17572.2 \pm 6255.8$}\\
		{FPFH} & {$459.3 \pm 341.8$}  & {$907.8 \pm 544.6$}\\
		{FPFH + ICP} & {$276.9 \pm 113.1$}  & {$ 550.5 \pm 102.7$}\\ \cline{1-3}
		{TSIM + MR} & {$668.2 \pm 1656.6$}  & {$1702.9 \pm 3411.3$}\\
		{TSIM + CPD} & {$3684.3 \pm 2615.5$}  & {$5008.4 \pm 3545.7$}\\
		{TSIM + FPFH} & {$401.6 \pm 188.0$}  & {$713.4 \pm 159.2$}\\
		{TSIM + FPFH + ICP} & {$224.2 \pm  103.6$}  & {$ 483.1 \pm 91.5$}\\
		{\bf Proposed method} & {$\bf 220.4 \pm 96.7$}  & {$\bf 471.6 \pm 87.7$}\\
		\hline
	\end{tabular}
\end{table}

\subsection{Correction of the IOS Stitching Errors}
\begin{figure*}
	\centering
	\subfloat{\includegraphics[width=.21\textwidth]{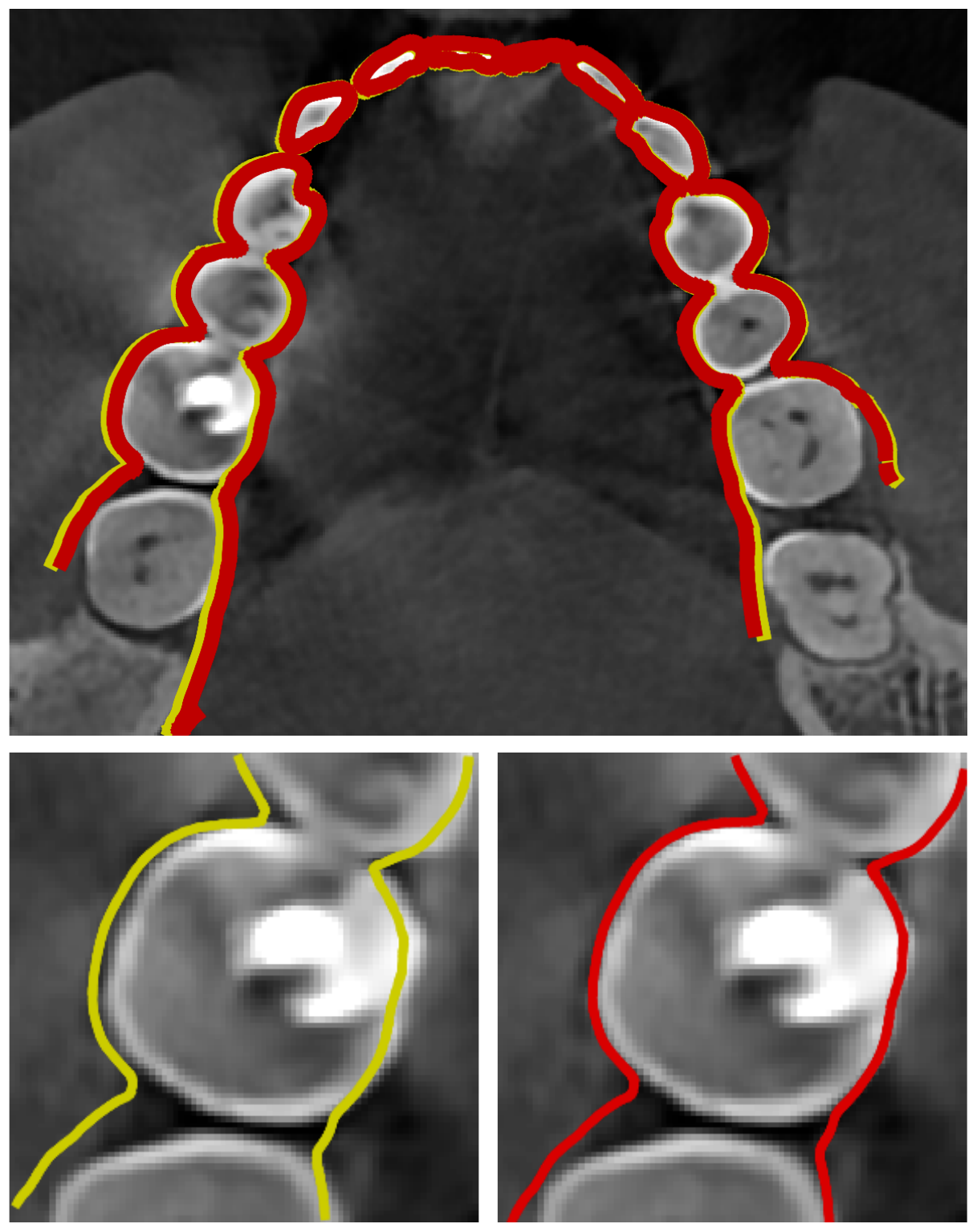}}~~~
	\subfloat{\includegraphics[width=.21\textwidth]{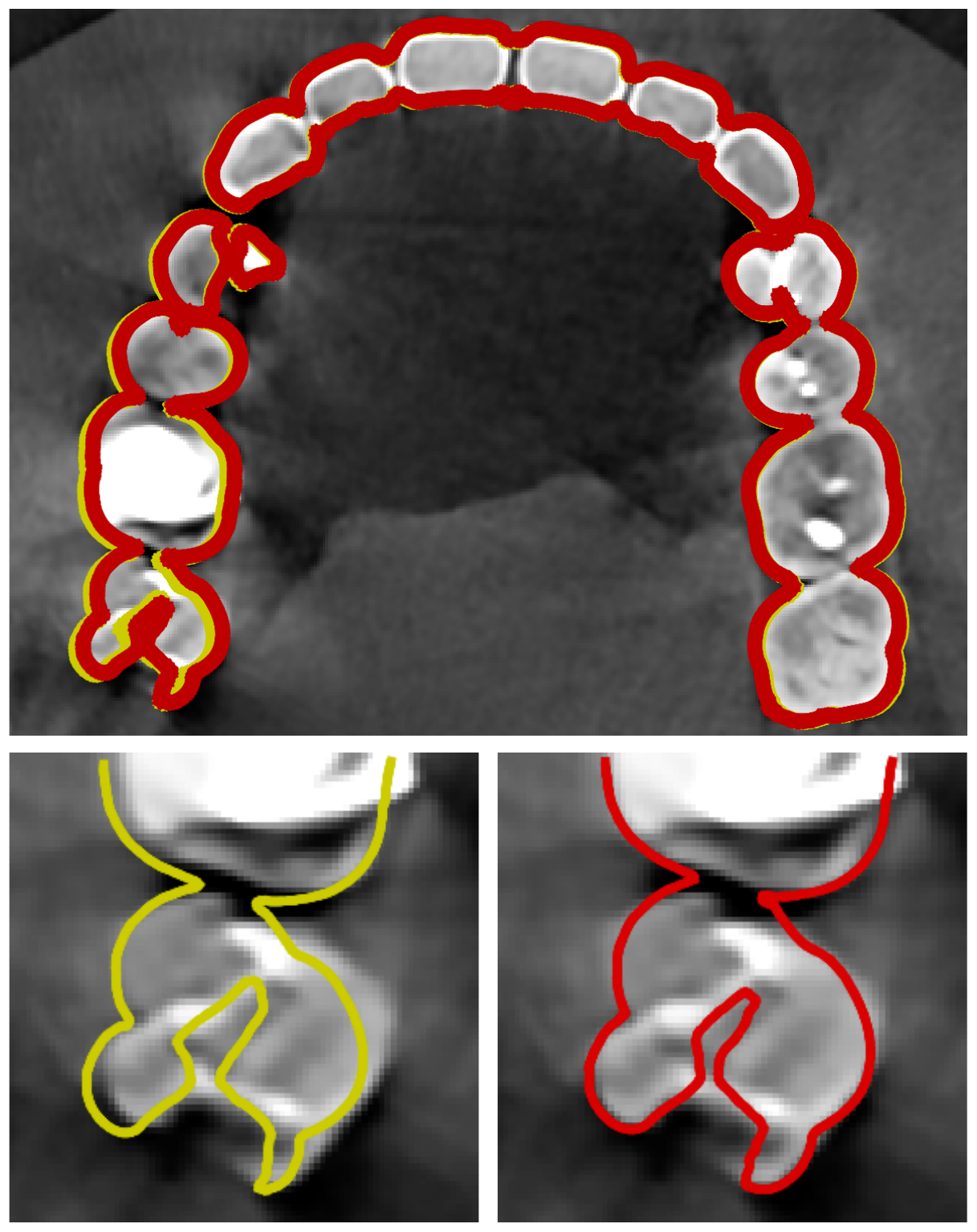}}~~~
	\subfloat{\includegraphics[width=.21\textwidth]{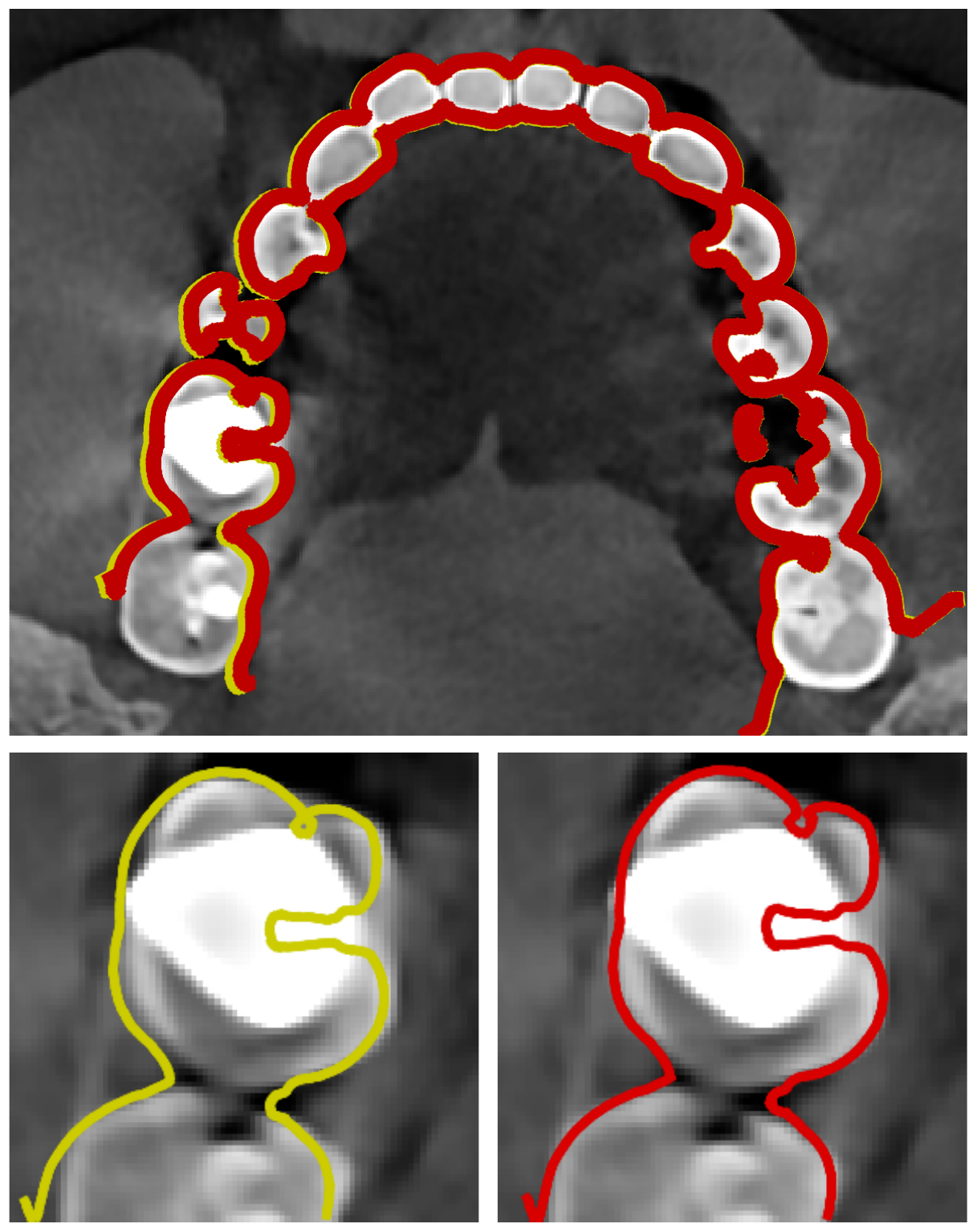}}~~~
	\subfloat{\includegraphics[width=.21\textwidth]{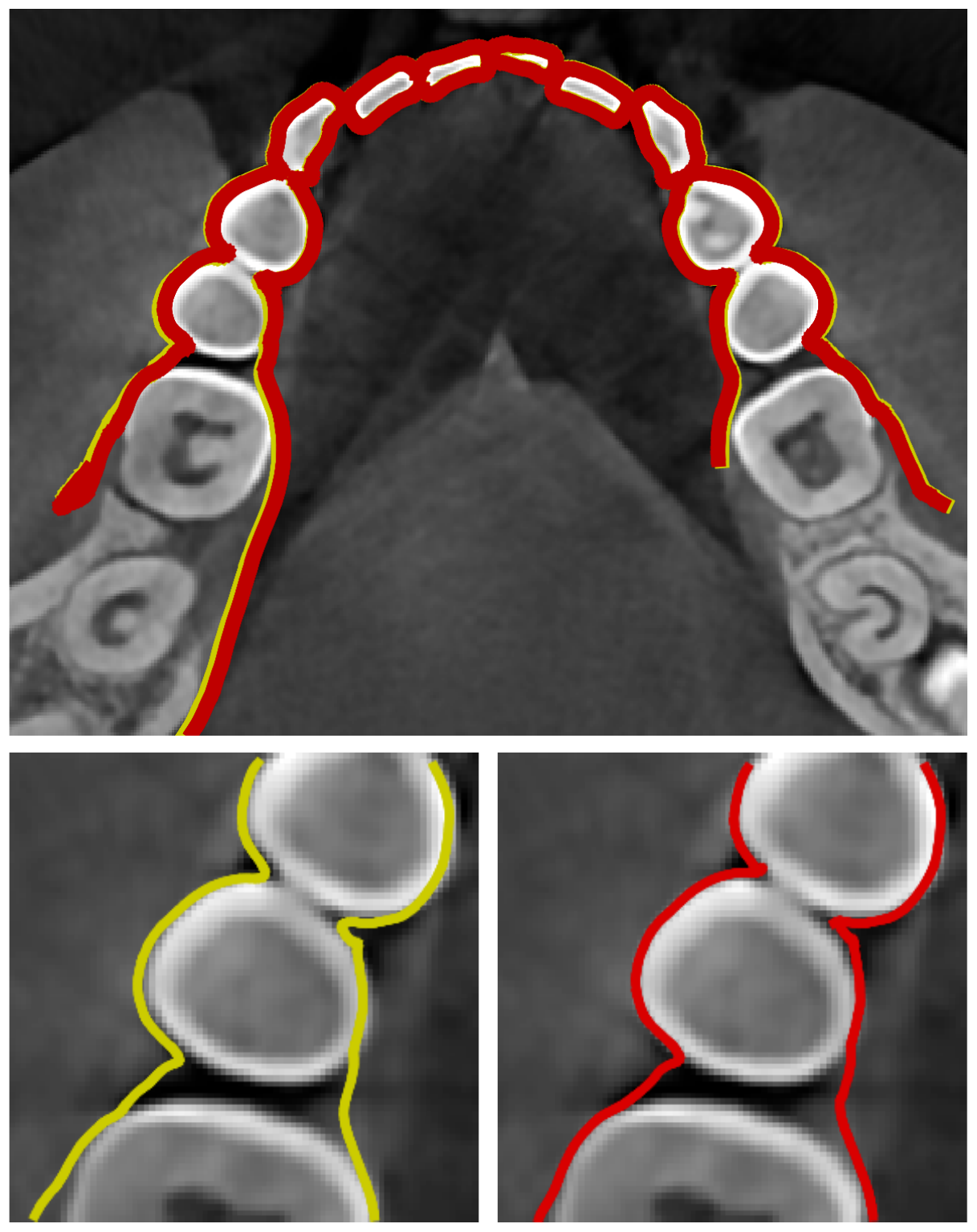}}
	\caption{Qualitative results before and after correction of four selected evaluation data. The yellow and red lines represent contours of the IOS models with the proposed registration and correction methods, respectively. The two contours almost overlap, but the differences appear at the end of the arches.}
	\label{fig:correction_results}
\end{figure*}

This subsection presents the result before and after the correction of distortions in IOS, which occur in the stitching process of locally scanned images. Table \ref{tbl:eval_cor} reports the correction results for the registration methods with TSIM that were used in the subsection above. All post-correction accuracies increased compared with the pre-correction accuracies. However, these correction results depend on the performances of registration methods. Each tooth of the IOS aligned with CBCT in the previous registration step is used as an initial guess to determine a corrective transformation. The locations of the IOS teeth should be as close as possible to the CBCT teeth, as the ICP may become stuck in local minima. Fig. \ref{fig:correction_results} presents the results of the proposed registration and correction methods. Due to accumulated stitching errors, the scanned arches tend to be narrower or wider than the actual arches. Thus, the registration results show that the full-arch IOS models slightly deviated at the end of the arches. In contrast, the corrected IOS models fit edges of the teeth in CBCT images. Additionally, we computed the $p$-values to show statistical significance between the proposed method and others.

\begin{table*}[]
	\footnotesize
	\centering
	{
	\caption{Results of Stitching Error Correction according to Registration Methods. \label{tbl:eval_cor}} \vskip 0.05in
	\begin{tabular}{cccccccc} \hline
		{\bf Method} & {\textbf{Landmark} ($\mu$m)} & {\textbf{Difference} ($\mu$m)} & {\bf \textit{p}-value} & {\textbf{Surface} ($\mu$m)} & {\textbf{Difference} ($\mu$m)} & {\bf \textit{p}-value} \\  \hline
		{TSIM + MR} & {$ 532.3 \pm 1694.4$}  & {$-135.9 \pm 37.8$} & {$<0.001$} & {$1492.8 \pm 3509.6$}  & {$-210.1 \pm 98.3$} & {$<0.001$}\\
		{TSIM + CPD} & {$3621.7 \pm 2709.2$}  & {$-62.6 \pm 93.7$} & {$<0.001$} & {$4935.5 \pm 3688.4$}  & {$-72.9 \pm 142.7 $} & {$<0.001$}\\
		{TSIM + FPFH} & {$144.3 \pm 91.9$}  & {$-257.3 \pm -96.1$} & {$<0.001$} & {$402.4 \pm 151.7$}  & {$ -311.0 \pm -7.5 $} & {$<0.001$}\\
		{TSIM + FPFH + ICP} & {$120.8 \pm 73.3$}  & {$-103.4 \pm -30.3$} & {$<0.01$} & {$323.3 \pm 118.4$}  & {$-160.1 \pm 27.2 $} & {$<0.001$} \\
		{\bf Proposed method} & {$\bf 112.4 \pm 73.1$}  & {$\bf -108.0 \pm -23.6$} & {$-$} & {$\bf 301.7 \pm 112.2$}  & {$\bf -169.9 \pm 24.5$} & {$-$}\\
		\hline
	\end{tabular}
	}
\end{table*}

\section{Discussion and Conclusion}

In this paper, we developed a fully automatic registration and correction technique that integrates two different imaging modalities (\textit{i.e.}, IOS and CBCT images) in one scene. The proposed method is intended not only to compensate CBCT-derived tooth surfaces with the high-resolution surfaces of IOS, but also to correct cumulative IOS stitching errors across the entire dental arch by referring to CBCT. The most important contribution of the proposed method is its registration accuracy at the level of clinical application, even with severe metal artifacts in CBCT. The accuracy is achieved by the use of TSIM-IOS and CBCT, which allow the minimization of the non-congruent points in CBCT and IOS data. The tooth-focused approach addresses the drawbacks of existing methods by achieving improved accuracy and fully automation. Moreover, this approach helps to correct full-arch digital impressions with distortion caused by stitching errors. A future study for further improvement is to show that the proposed method is effective in various clinical cases with partial edentulousness or dental braces. It was difficult to gather full-arch intraoral impression data because intraoral scanners are rarely used for full-arch scanning due to accumulated stitching errors. We leave another future research for more precise validation of the IOS correction results (\textit{e.g.}, cadaver study), as the proposed method evaluated the IOS accuracy using CBCT images.

The fusion of the CBCT images and IOS models provides high-resolution crown surface even in the presence of serious metal-related artifacts in the CBCT images. Metal artifact reduction (MAR) in dental CBCT is known to be the most difficult and important issue. By avoiding the challenging problem of MAR with the help of IOS, the merged image may be used for occlusal analysis. The proposed multimodal data integration system can provide a jaw-tooth-gingiva composite model, which is a basic tool in digital dentistry workflow. Thus, it may be used to produce a surgical wafer for orthognathic surgical planning and orthodontic mini-screw guide to reduce failure by minimizing root contact. Furthermore, because the jaw-tooth-gingiva model is componentized with jaw bones, individual teeth, and soft tissues (gingiva and palate), it is useful in terms of versatility and practicality in various dental clinical simulation and evaluation.

The proposed method can eliminate the hassle of traditional dental prosthetic treatments that are labor intensive, costly, require at least two individual visits, and require temporary prosthesis to be worn until the final crown is in place. Moreover, if the final crown made in the dental laboratory does not fit properly at the second visit, the patient and dentist will have to repeat the previous operation, and the laboratory may have to redesign the restoration prosthesis. Note that the proposed integration of dental CBCT and IOS data can provide an alternative to traditional impressions, thereby reducing the time-consuming laboratory procedure of manually editing individual teeth using a computer-aided interface.

\section*{Acknowledgment}
This research was supported by a grant of the Korea Health Technology R\&D Project through the Korea Health Industry Development Institute (KHIDI), funded by the Ministry of Health \& Welfare, Republic of Korea (grant number : HI20C0127). We would like to express our deepest gratitude to HDXWILL which shares dataset.
\bibliographystyle{IEEEtran}
\bibliography{manuscript.bib}

\clearpage

\begin{figure*}
	\centering
	\includegraphics[width=.95\textwidth]{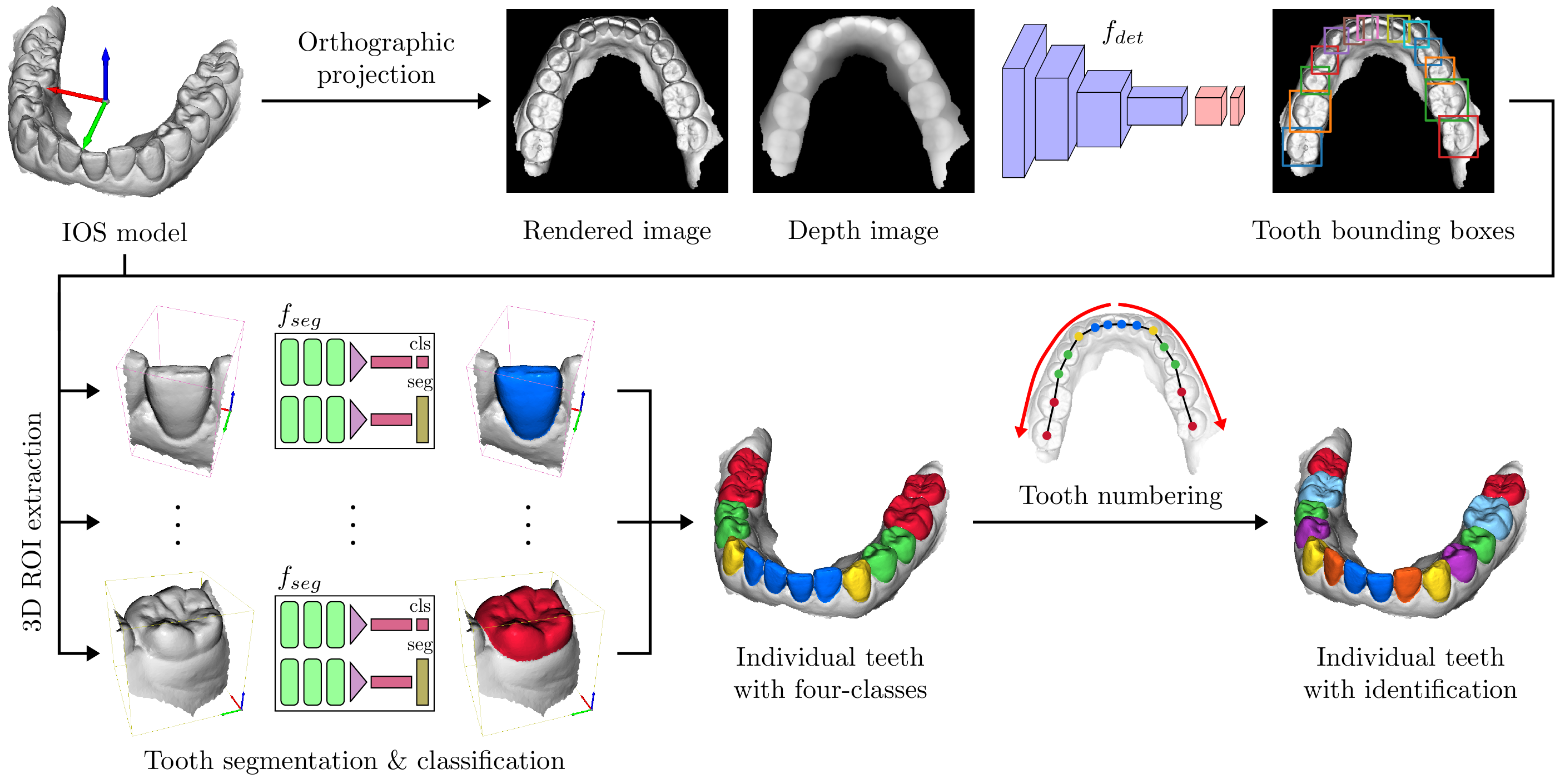}
	\caption{Flowchart of the proposed method for individual tooth segmentation and identification in IOS. The tooth highlighted-feature 2D images of a given 3D IOS model are used to determine the 3D tooth ROIs. Next, individual tooth segmentation and identification are performed using the obtained tooth ROIs. Here, the teeth are classified into four tooth types and sequentially assigned the numbers from incisors to the molars.}
	\label{fig:ios_seg}
\end{figure*}

\appendices
\section{Deep learning based individual tooth segmentation and identification in IOS} \label{app:sec1}

This section explains the deep learning-based method for the segmentation of individual teeth and the identification of tooth classes from IOS data. Let $X$ denote the point cloud of a full-arch surface model scanned by IOS. The goal of the segmentation is to decompose $X$ into individual teeth ($X_{t_1}\cup \cdots \cup X_{t_J}$) and the rest including gingiva ($X_{\mbox{\scriptsize gingiva}} $):

\begin{equation}\label{decompose-tooth}
X=\underbrace{X_{t_1}\cup \cdots \cup X_{t_J}}_{X_{\mbox{\scriptsize teeth}}}\cup X_{\mbox{\scriptsize gingiva}},
\end{equation}
where $J$ is the number of teeth in the IOS model (\textit{i.e.}, $J\le 16$) and each $X_j$ represents the tooth crown with the code $t_j$ according to the universal notation system \cite{nelson2014wheeler}.

The proposed method consists of three steps: i) tooth feature-highlighted 2D image generation from IOS, ii) 2D bounding box detection and 3D tooth region of interest (ROI) extraction, and iii) 3D individual tooth segmentation and identification. The workflow of the proposed method is illustrated in Fig. \ref{fig:ios_seg}.

\begin{figure}
	\centering
	\subfloat[]{\includegraphics[width=0.35\textwidth]{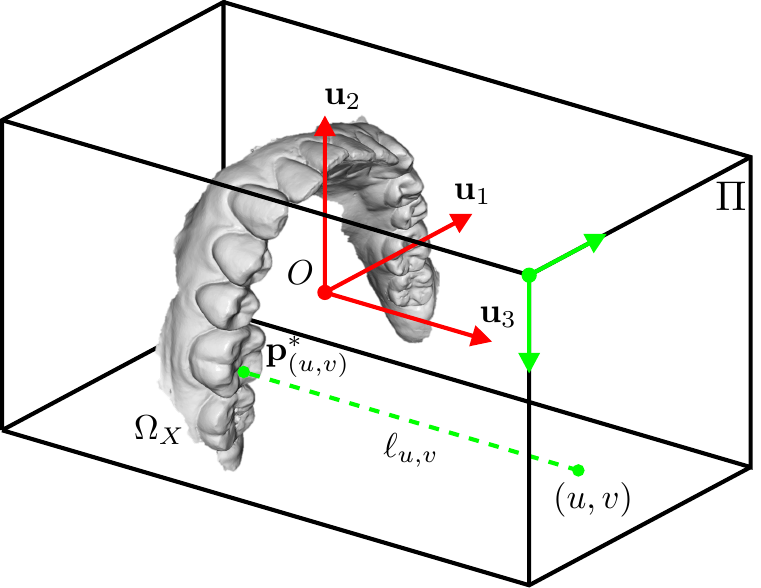}}
	\hfill
	\subfloat[]{\includegraphics[width=0.2\textwidth]{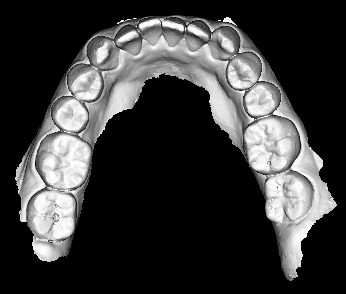}}~~~~
	\subfloat[]{\includegraphics[width=0.2\textwidth]{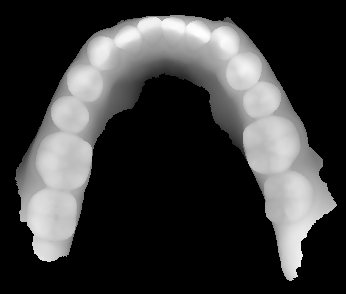}}
	\caption{Generation of 2D rendered images with a lighting effect and depth images from $\Omega_X$ by orthographic projection. (a) Alignment of $X$ in a new coordinate system with three axes $\u_1,\u_2,\u_3$. The generated images are defined on $\Pi$ determined by $\u_1,\u_2,\u_3$. (b) Rendered image $\mathcal{I}_r$, and (c) depth image $\mathcal{I}_d$ where the grayscale stands for the depth information.}
	\label{fig:ios_step1}
\end{figure}

\subsection{Tooth feature-highlighted 2D image generation}
Let $\Omega_X$ be a point-set surface of the point cloud data $X$. From the surface $\Omega_X$, we can generate tooth feature-highlighted 2D images; 2D rendered images with lighting effect (denoted by $\mathcal{I}_{r}$) and depth images (denoted by $\mathcal{I}_d$) utilizing the orthographic projection technique \cite{lengyel2011mathematics}. These two images will be used for the detection of bounding boxes containing individual teeth.

To generate $\mathcal{I}_r$ and $\mathcal{I}_d$, we first need to align $X$ in a new coordinate system with three axes, namely, $\u_1$, $\u_2$, and $\u_3$ (Fig. \ref{fig:ios_step1}a), which are roughly horizontal, sagittal, and vertical directions, respectively. The origin of the coordinate system is selected as $\bar{X}=\f{1}{N}\sum_{i=1}^N \x_i$. We apply the principal component analysis (PCA) to obtain three principal bases $\{\mathbf{pc}_1,\mathbf{pc}_2,\mathbf{pc}_3\}$ for $X-\bar{X} \coloneqq \{\x-\bar{X}:\x \in X\}$. The three coordinate directions $\{\u_1,\u_2,\u_3\}$ are then chosen by

\begin{align}
	&\u_2 =
	\begin{cases}
		{\mathbf{pc}_2}&{\text{if}~\left< \mathbf{pc}_2, \sum_{\x \in X}\frac{\x-\bar{X}}{\|\x-\bar{X}\|} \right>} \geq 0\\
		{-\mathbf{pc}_2}&{\text{otherwise}}
	\end{cases},
	\\
	&\u_3 =
	\begin{cases}
		{\mathbf{pc}_3}&{\text{if}~\left< \mathbf{pc}_3, \sum_{\x \in X}\mathbf{n}_\x \right>} \geq 0\\
		{-\mathbf{pc}_3}&{\text{otherwise}}
	\end{cases},
	\\
	&\u_1 = \u_2 \times \u_3,
\end{align}
where $\n_{\x}$ is a normal vector at a point $\x \in X$.

We will now explain how to generate the 2D tooth feature-highlighted images $\mathcal{I}_r$ and $\mathcal{I}_d$. Without loss of generality, we may assume that $\bar{X}=0$ and $\u_1=(1,0,0),\u_2=(0,1,0),\u_3=(0,0,1)$. As shown in Fig. \ref{fig:ios_step1}a, the rendered 2D image $\mathcal{I}_r(u,v)$ is defined on the occlusal plane $\Pi$:

\begin{equation}
	\Pi = \left\{ s((u, -v, 0) + \a) : 1 \leq u \leq N_1,~ 1 \leq v \leq N_2 \right\},
\end{equation}
where $\a \coloneqq (-\frac{N_1+1}{2},\frac{N_2+1}{2},\max\{\|\mathbf{p}\| : \mathbf{p} \in \Omega_X\})$ is a translation vector and $s$ is the pixel spacing. Here, $s$ and $N_1 \times N_2$ (\textit{e.g.}, $s=0.2$ and $N_1 \times N_2 = 400 \times 400$) are related to the image resolution and field of view, respectively. To be precise, $\mathcal{I}_r$ is given by

\begin{equation}
	\mathcal{I}_r(u,v) =
	\begin{cases}
		{\max\{ \langle \mathbf{n}_{\mathbf{p}^*_{u,v}},\u_3 \rangle,0\}} & {\text{if}~\ell_{u,v} \cap \Omega_X \neq \emptyset} \\
		{0} & {\text{otherwise}}
	\end{cases},
\end{equation}
where $\ell_{u,v}$ is the line passing through $s((u,-v,0)+\a)$ with the direction $\u_3$, $\n_\mathbf{p}$ is a unit normal vector at $\mathbf{p}$, and $\mathbf{p}^*_{u,v}$ is a point lying on the tooth surface $\Omega_X$ given by

\begin{equation}
	\mathbf{p}^*_{u,v} \coloneqq \mbox{argmax}\{\langle \mathbf{p},\u_3 \rangle:\mathbf{p} \in \ell_{u,v} \cap \Omega_X\}.
\end{equation}

The depth image $\mathcal{I}_d$ is given by

\begin{equation}
	\mathcal{I}_d(u,v)=
	\begin{cases}
		{1-\frac{\langle \mathbf{p}^*_{u,v},\u_3 \rangle - z^*_{\mbox{\tiny min}}}{z^*_{\mbox{\tiny max}}-z^*_{\mbox{\tiny min}}}} & {\text{if}~\ell_{u,v} \cap \Omega_X \neq \emptyset} \\
		{0} & {\text{otherwise}}
	\end{cases},
\end{equation}
where $z^*_{\mbox{\tiny min}} \coloneqq \min\{\langle \mathbf{p}^*_{u,v},\u_3 \rangle : (u,v) \in N_1 \times N_2 \}$ and $z^*_{\mbox{\tiny max}} \coloneqq \max\{\langle \mathbf{p}^*_{u,v},\u_3 \rangle : (u,v) \in N_1 \times N_2 \}$. As shown in Figs. \ref{fig:ios_step1}b and \ref{fig:ios_step1}c, the depth values of the tooth crowns are distinct because the tooth positions protrude forward compared to the gingiva and other tissues. While the rendered image contains geometric features of the surface by the lighting and shading effects, the depth image provides the tooth reliability information by expressing the relative distance.

\subsection{Tooth bounding box detection and 3D tooth ROI extraction}

We use a convolutional network-based method\cite{redmon2016you} to obtain a tooth detection map:

\begin{equation}\label{f_det}
  f_{det} : (\mathcal{I}_r, \mathcal{I}_d)  \mapsto \{ \b_{1}, \cdots, \b_{J} \},
\end{equation}
where $\{ \b_{1}, \cdots, \b_{J} \}$ denotes a set of vectors associated with the 2D bounding boxes corresponding to individual teeth. Each $\mathbf{b}_{j} = (u_j^1,v_j^1,u_j^2,v_j^2)$ should contain a single tooth and can be uniquely determined by a bounding box coordinates with the left top corner $(u_j^1,v_j^1)$ and right down corner $(u_j^2,v_j^2)$.

The individual tooth ROIs $\{X_{roj_1},\cdots,X_{roj_J}\}$ are determined by the bounding box components $\{\b_{1},\cdots,\b_{J}\}$:

\begin{equation}
	X_{{roi}_{j}} = X \cap \mbox{roi}_{j},
\end{equation}
where $\mbox{roi}_{j} = \left[su_j^1,su_j^2 \right) \times \left[ -sv_j^2,-sv_j^1\right) \times \R $.

\subsection{3D individual tooth segmentation and identification}

\begin{figure}[ht]
	\centering
	\subfloat[]{\includegraphics[width=0.22\textwidth]{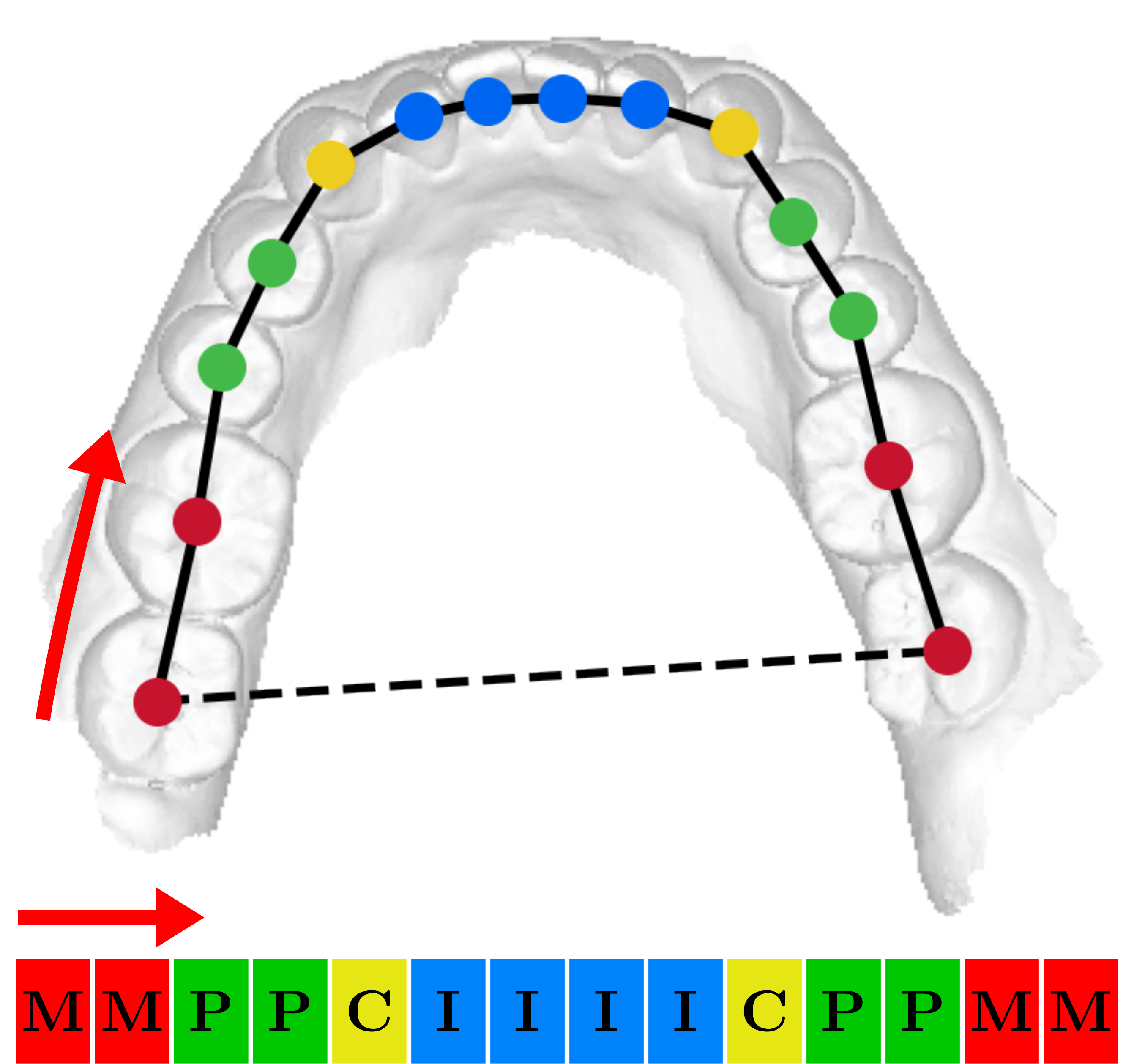}}~~~~~
	\subfloat[]{\includegraphics[width=0.22\textwidth]{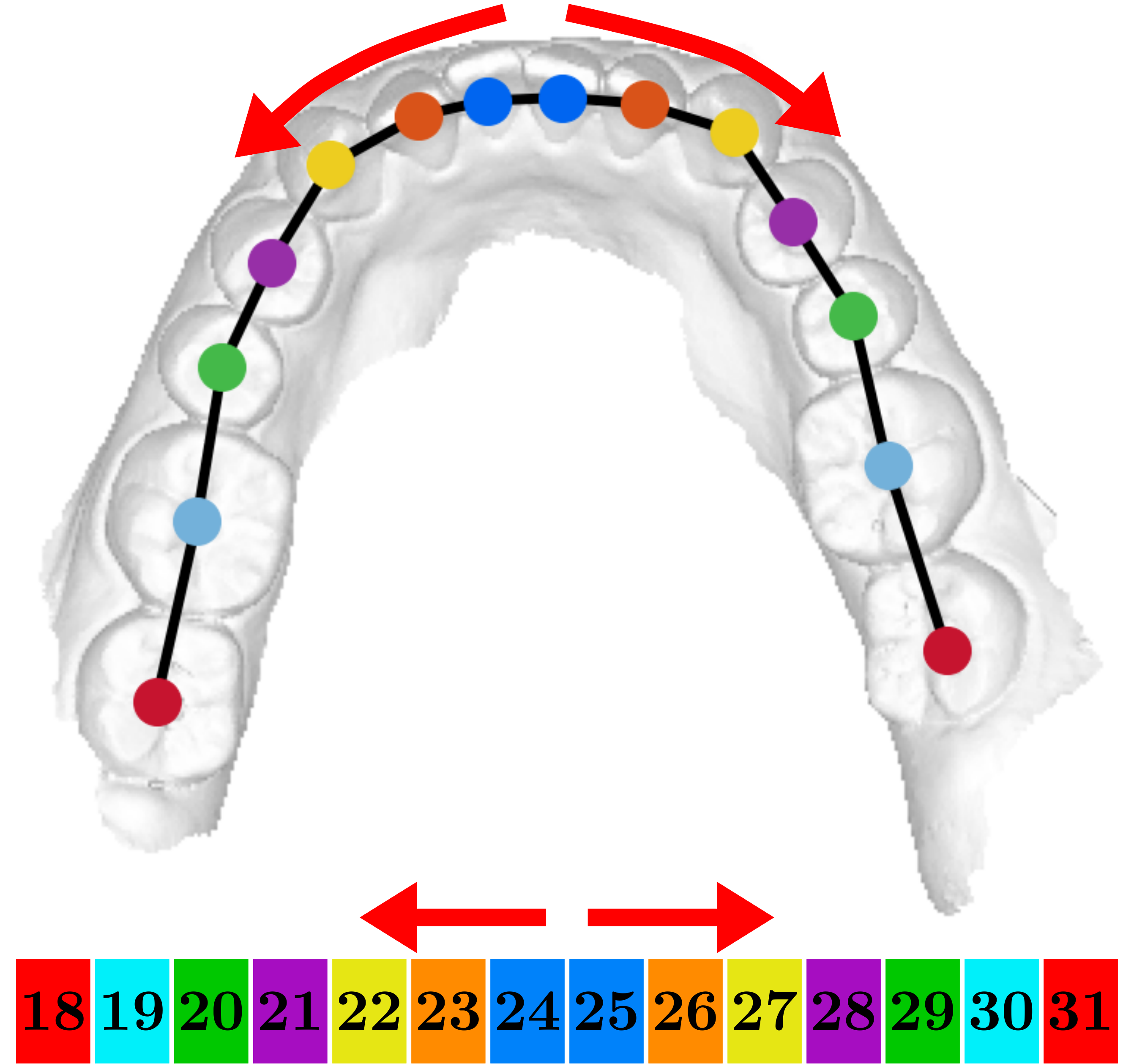}}
	\caption{Tooth identification from four-classes tooth classification. (a) Arrangement of the teeth with four types through a concave hull for the tooth center positions. (b) Assignment of codes to teeth according to the universal dental notation. The teeth are numbered sequentially from the inside (left and right incisors) to the outside (left and right molars).}
	\label{fig:ios_step3}
\end{figure}

In this step of tooth segmentation and identification, $(X_j,t_j)$ is obtained for each 3D tooth ROI $X_{roi_j}$, where $X_{j} \subset X_{roi_j}$ is the segmented tooth and $t_j$ is the tooth code corresponding to $X_{j}$. First, using a graph-based multi-layer perceptron \cite{wang2019dynamic} for point cloud segmentation, we obtain a tooth segmentation map:

\begin{equation}
	f_{seg} : X_{roi_j} \mapsto (X_j,c_j),
\end{equation}
where $c_j$ represents one of four tooth classes: incisors, canines, premolars, and molars. Here, end-to-end prediction for tooth identification is difficult because the geometric shapes of the adjacent teeth are similar \cite{jang2021fully}. Therefore, it is reasonable to classify teeth into four types according to tooth morphology. To accomplish tooth identification, we use the center point $(u_j,v_j) \coloneqq \frac{1}{2}({u_j^1+u_j^2},-v_j^2-v_j^1)$ of the bounding box component $\b_j$. As shown in Fig. \ref{fig:ios_step3}a, a concave hull is obtained from $\{(u_j,v_j) : j=1,\cdots,J\}$ by applying the ball pivoting algorithm \cite{bernardini1999ball}, which is a shape reconstruction method related to the alpha shape \cite{edelsbrunner1983shape}. Upon removing the longest side in the concave hull, the remaining piecewise linear curve represents the connection relationship between adjacent teeth. We can then list the values of $c_j$'s according to their concatenation order. Next, as shown in Fig. \ref{fig:ios_step3}b, each tooth is sequentially assigned a code number $t_j$ suitable for the type $c_j$. When the identification process is complete, we refer to $(X_j,t_j)$ as $X_{t_j}$.

\section{Fast Point feature histogram}\label{app:sec2}

To address the limitations of ICP-based methods, which often fall into local minima during registration, Rusu \textit{et al.} \cite{rusu2009fast} developed a global registration method called Fast Point Feature Histograms (FPFH) for feature-based matching between two point clouds. FPFH is designed to find pairs of points with similar geometric features between two point clouds.

We now explain $\mbox{FPFH}(\x)$, whose definition is somewhat complicated. $\mbox{FPFH}(\x)$ is a multi-dimensional vector representing local geometric features of a point cloud $X$ at $\x$ and relevant information from its surrounding neighborhood points over $X$. Furthermore, $\mbox{FPFH}$ is defined in a discriminatory manner to reliably discriminate regional geometric features. 

Let $NN_{k}(\x)$ denote the $k$-nearest neighborhood of $\x$ over the point cloud $X$. To be precise, $NN_{k}(\x)$ is the set of $k$ points such that, for all $\tilde{\x} \in X \setminus NN_{k}(\x)$,

\begin{equation}\label{eq:nn}
	\max \left\{ \|\x'-\x\| : \x' \in NN_{k}(\x) \right\} \le \|\tilde{\x}-\x\|.
\end{equation} 
Let $\n_\x$ denote the unit normal vector of the point cloud $X$ at $\x$. The $\mbox{FPFH}(\x)$ is based on the angular variations of the normals on $NN_{k}(\x)$. For $\x'\in NN_{k}(\x)\setminus \{\x\}$, we compute the following three angles $\{\rho(\x,\x'), \phi(\x,\x'), \theta(\x,\x')\}$, which are designed to be symmetric \cite{wahl2003surflet}:

\begin{align}
	& \rho(\x,\x') = \cos^{-1} \left( \langle \u,\frac{\x_t-\x_s}{\|\x_t-\x_s\|}\rangle \right), \\
	& \phi(\x,\x') = \cos^{-1} \left( \langle \v, \n_{\x_t} \rangle \right), \\
	& \theta(\x,\x') =
	\begin{cases}
		{\cos^{-1}\left( \langle \u, \n_{\x_t} \rangle \right)} & \text{if }  \langle \w, \n_{\x_t} \rangle \geq 0 \\
		{\cos^{-1}\left( \langle \u, \n_{\x_t} \rangle \right)+\pi} & {\text{otherwise}}
	\end{cases},
\end{align}
where $(\x_s,\x_t)$ is either $(\x,\x')$ or $(\x',\x)$, which is determined by

\begin{equation}\label{xsxt}
	(\x_s,\x_t)=
	\begin{cases}
		{(\x,\x')}&{\text{if } \langle \n_\x, \x'-\x \rangle \leq \langle \n_{\x'}, \x-\x' \rangle} \\
		{(\x',\x)}&{\text{otherwise}}
	\end{cases},
\end{equation}
and the triple $\u,\v,\w$ is the Darboux frame defined by
$$
\u=\n_{\x_s},~~~ \v=\frac{\x_t-\x_s}{\|\x_t-\x_s\|}\times \u,~~~\w=\u\times \v.
$$
Note that the use of $(\x_s,\x_t)$ in Eq. \eqref{xsxt} is needed in order that the three angles $\rho$, $\phi$, and $\theta$ have the symmetric property. These three angles provide consistent geometric features that represent the difference between $\n_\x$ and $\n_{\x'}$.

Next, we define a simplified point feature histogram (SPFH):

\begin{equation}
	\mbox{SPFH}(\x) = \left(\varrho(\x),\varphi(\x),\vartheta(\x) \right) \in \R^{11} \times \R^{11} \times \R^{11},
\end{equation}
where

\begin{align}
	&\varrho(\x) = \frac{1}{k} \sum_{\x' \in NN_{k}(\x)\setminus \{\x\}} \hbar(\rho(\x,\x')), \\
	&\varphi(\x) = \frac{1}{k} \sum_{\x' \in NN_{k}(\x)\setminus \{\x\}} \hbar(\phi(\x,\x')), \\
	&\vartheta(\x) = \frac{1}{k} \sum_{\x' \in NN_{k}(\x)\setminus \{\x\}} \hbar(\frac{1}{2} \theta(\x,\x')).
\end{align}
Here, $\hbar : [0,\pi) \mapsto \R^{11}$ is the map defined by

\begin{equation} 
	\hbar(s)=(\hbar_1(s),\cdots,\hbar_{11}(s)),
\end{equation}
where

\begin{equation}
	\hbar_j(s) =
	\begin{cases}
		{1}&{\text{if}~s \in [\frac{j-1}{11}\pi,\frac{j}{11}\pi)} \\
		{0}&{\text{otherwise}}
	\end{cases}.
\end{equation}
The vector-valued function $\hbar$ is used to increase the ability to discriminate differences between different local geometric features. Then, $\mbox{FPFH}(\x)$ is defined as weighted sum of $\mbox{SPFH}(\x')$ over  $NN_{k}(\x)$:

\begin{equation}
	\mbox{FPFH}(\x) = \mbox{SPFH}(\x) + \frac{1}{k}\sum_{\x' \in NN_{k}(\x)\setminus \{\x\}} \frac{\mbox{SPFH}(\x')}{1+{\|\x-\x'\|}}.
\end{equation}
Here, the weight $1/(1+{\|\x-\x'\|})$ depends on the center point $\x$ and its neighbor $\x'\in NN_{k}(\x)\setminus \{\x\}$.

%
%

%
%
%

\end{document}